\documentclass[epj]{svjour}
\usepackage{graphicx,cite}

\def\eqref#1{(\ref{#1})}

\newcommand{\power}[3]{\left({{#1} \over {#2}}\right)^{#3}} 

\newcommand{\inverse}[1]{ {1  \over {#1}} }

\newcommand{\twid}{\sim}

\newcommand{\paren}[1]{\left( #1 \right)}
\newcommand{\square}[1]{\left[ #1 \right]}

\newcommand{\casesbracketsshortii}[4]
{\left\{
\begin{array}{ll}
#1\ & (#2) \\  #3\ & (#4) 
\end{array}%
\right.
}

\newcommand{\casesbracketsshortiii}[6]{
\left\{
\begin{array}{lll} 
#1\  & (#2) \\  #3\ & (#4) \\  #5\ &(#6)   
\end{array}%
\right.
}

\newcommand{\gap}{\hspace{.4in}}

\newcommand{\blank}{\ \\}

\newcommand{\gt}{\rightarrow}

\newcommand{\period}{\ \ .}
\newcommand{\comma}{\ ,\ }

\newcommand{\lsim}{\,\stackrel{<}{\scriptstyle \sim}\,}

\newcommand{\ignore}[1]{}

{\end{list}}

{\end{list}}

{\ \\ XXXXX \hfill XXXXX #1 XXXXX\begin{equation}\label{#1}}%
{\end{equation}}

{\ \\ XXXXX \hfill XXXXX #1 XXXXX\begin{eqnarray}\label{#1}}%
{\end{eqnarray}}

\newenvironment{bequation}[1]%
{\begin{equation}\label{#1}}%
{\end{equation}}

\newenvironment{beqnarray}[1]%
{\begin{eqnarray}\label{#1}}%
{\end{eqnarray}}

\newcommand{\drop}{\nonumber \\}

\newcommand{\ddrop}{\drop\drop}

\newcommand{\ie}{i.\,e.~}
\newcommand{\eg}{e.\,g.\@ }

{\begin{center} \section*{#1} \end{center}}

{\end{figure}} 

{\end{figure}}

{\end{figure}}

\newenvironment{eq}[1]%
{\begin{bequation}{#1}}{\end{bequation}}

\newenvironment{eqarray}[1]%
{\begin{beqnarray}{#1}}{\end{beqnarray}}

\newcommand{\smax}{s^{\rm max}}

\newcommand{\staumax}{s_{\tau}^{\rm max}}

\newcommand{\Mtot}{M^{\rm tot}}

\newcommand{\omegatwiddle}{\widetilde {\omega}}

\newcommand{\tauN}{\tau_N}

\newcommand{\phisurf}{\phi_{\rm surf}}
\newcommand{\phibulk}{\phi_{\rm bulk}}
\newcommand{\phistar}{\phi^{*}}
\newcommand{\Zsurf}{Z_{\rm surf}}
\newcommand{\Zbulk}{Z_{\rm bulk}}

\newcommand{\kT}{k_B T}

\newcommand{\Gammabound}{\Gamma_{\rm bound}}
\newcommand{\gammabound}{\gamma_{\rm bound}}

\newcommand{\Rtotal}{{\cal R}_{\rm total}}

\newcommand{\omegatau}{\omega_{\tau}}

\newcommand{\tsatchem}{t_{\rm sat}^{\rm chem}}
\newcommand{\tsatphys}{t_{\rm sat}^{\rm phys}}
\newcommand{\tfinalphys}{t_{\rm final}^{\rm phys}}

\newcommand{\taubulk}{\tau_{\rm bulk}} 

\newcommand{\RF}{R_F}

\newcommand{\ta}{t_a}

\newcommand{\ncont}{n_{\rm cont}}

\newcommand{\rhosuper}{\rho_{\rm super}}
\newcommand{\Gammaboundinf}{\Gamma_{\rm bound}^{\infty}}
\newcommand{\Gammainf}{\Gamma^{\infty}}

\newcommand{\lsep}{l_{\rm sep}}

\newcommand{\Zloop}{Z_{\rm loop}}
\newcommand{\Zcont}{Z_{\rm cont}}

\newcommand{\Nsite}{N_{\rm site}}

\newcommand{\toverlap}{t_{\rm overlap}}

\newcommand{\Dtau}{D_\tau} 

\newcommand{\tauads}{\tau_{\rm ads}}
\newcommand{\Ncollisions}{N_{\rm coll}}

\newcommand{\nth}{^{\rm th}}

\newcommand{\Peq}{P^{\rm eq}}
\newcommand{\fbar}{\bar{f}}

\begin{document}
%

\title{Irreversible Adsorption from Dilute Polymer Solutions}

\author{
Ben O'Shaughnessy \inst{1} \thanks{\email{bo8@columbia.edu}}
\and 
Dimitrios Vavylonis\inst{1,2} \thanks{\email{dv35@columbia.edu}}
}                     
%
%
\institute{
Department of Chemical Engineering, Columbia University, 
500 West 120th Street, New York, NY 10027, USA
 \and
Department of Physics, Columbia University, 
538 West 120th Street, New York, NY 10027, USA
}

\date{}
%


\abstract{We study irreversible polymer adsorption from dilute
solutions theoretically. Universal features of the resultant
non-equilibrium layers are predicted. Two broad cases are considered,
distinguished by the magnitude of the local monomer-surface sticking
rate $Q$: chemisorption (very small $Q$) and physisorption (large
$Q$).  Early stages of layer formation entail single chain adsorption.
While single chain physisorption times $\tauads$ are typically micro
to milli-seconds, for chemisorbing chains of $N$ units we find
experimentally accessible times $\tauads = Q^{-1} N^{3/5}$, ranging
from seconds to hours. We establish 3 chemisorption universality
classes, determined by a critical contact exponent: zipping,
accelerated zipping and homogeneous collapse. For dilute solutions,
the mechanism is accelerated zipping: zipping propagates outwards from
the first attachment, accelerated by occasional formation of large
loops which nucleate further zipping.  This leads to a transient
distribution $\omega(s) \twid s^{-7/5}$ of loop lengths $s$ up to a
maximum size $\smax \approx (Q t)^{5/3}$ after time $t$.  By times of
order $\tauads$ the entire chain is adsorbed.  The outcome of the
single chain adsorption episode is a monolayer of fully collapsed
chains.  Having only a few vacant sites to adsorb onto, late arriving
chains form a diffuse outer layer.  In a simple picture we find for
both chemisorption and physisorption a final loop distribution
$\Omega(s) \twid s^{-11/5}$ and density profile $c(z) \twid z^{-4/3}$
whose forms are the same as for equilibrium layers.  In contrast to
equilibrium layers, however, the statistical properties of a given
chain depend on its adsorption time; the outer layer contains many
classes of chain, each characterized by a different fraction of
adsorbed monomers $f$.  Consistent with strong physisorption
experiments, we find the $f$ values follow a distribution $P(f) \twid
f^{-4/5}$.  } 



%
\maketitle
%



\section{Introduction}

\begin{figure}
\centering
\resizebox{0.8\columnwidth}{!}{%
  \includegraphics{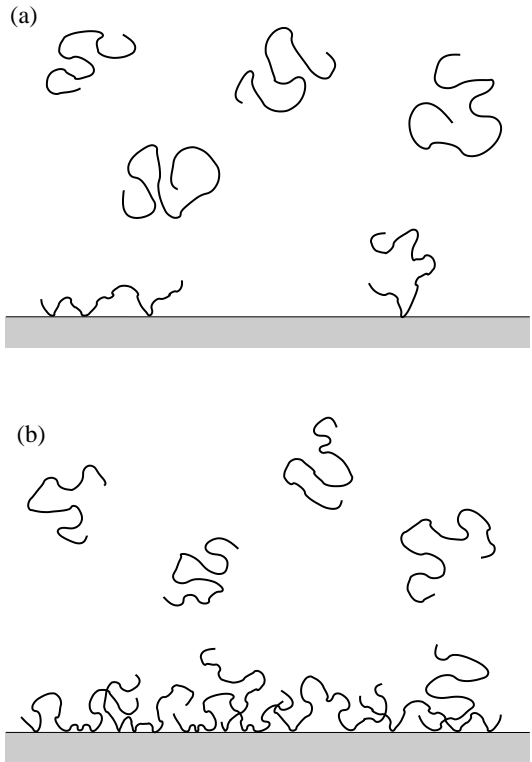}
}
\caption{The situation studied in this paper: polymer chains adsorbing
onto a surface from a dilute polymer solution.  We consider situations
where monomer-surface bonds develop irreversibly, due either to
chemical bonding or strong physical interactions.  (a) Early stages of
layer formation.  The surface is almost empty and the first chains to
arrive adsorb on the surface without interference from others. (b) At
longer times a polymer layer of strongly interacting chains develops
and chain densities on the surface become much higher than those in
the bulk.}
\label{chemiphysi_scheme}      
\end{figure}

Polymer layer formation is unavoidable when even weakly attractive
surfaces come into contact with a polymer solution
\cite{gennes:pols_interface_review,fleer:pol_iface_book}, see
fig. \ref{chemiphysi_scheme}.  Even for extremely dilute polymer
solutions, polymer layers develop with densities which may be many
orders of magnitude larger than the bulk polymer concentration
\cite{bouchauddaoud:ads_pol_concentration_effects}.  This is due to
the fact that by giving up their bulk translational entropy, which
costs a free energy of only $kT$, chains achieve an energy advantage
proportional to the number of monomers per chain, $N$, for large
enough $N$ values.  The topologically complex interfacial layers
contain both surface-bound segments and large loops and tails
extending into the bulk (see fig. \ref{chemiphysi_scheme}).  In
principle, given enough time, adsorbed polymers are able to explore
all accessible states
\cite{gennes:ads_pol_dynamics_toyota,gennes:ads_pol_dynamics_new_trends,%
semenovjoanny:kinetics_adsorption_rouse} and the layer attains
equilibrium.  Equilibrium layers have been the focus of a large body
of experimental
\cite{fleer:pol_iface_book,lee:neutron_reflectivity_adsorbed_pol,%
auvraycotton:neutron_scattering_adsorbed_pol,kawaguchitakahashi:ads_pol_ellipsometry},
theoretical
\cite{gennes:pols_interface_review,fleer:pol_iface_book,bouchauddaoud:ads_pol_concentration_effects,%
gennes:self_similar_profile,gennespincus:proximal_exponent,%
eisenriegler:proximal_exponent,gennes:ads_pol_loops,aubouy:scaling_flat_layers,%
semenovjoanny:loops_tails_europhys,%
scheutjensfleer:mean_field_adsorbed_pol_1,scheutjensfleer:mean_field_adsorbed_pol_2,%
jonesrichmont:mean_field_adsorbed_pol,semenov:loops_tails_macro,%
johner:pol_ads_interpretation_numerical_results} and numerical
\cite{lai:pol_ads_statics_dynamics_monte_carlo,%
zajacchakrabarti:pol_ads_statics_dynamics_monte_carlo,dejoannis:polads_monte_carlo}
work.

In a great many cases, however, the surface sticking energy per
monomer $\epsilon$ exceeds $\kT$.  Available experimental evidence
suggests that desorption processes and relaxation kinetics within the
layer are then sharply slowed down
\cite{pefferkon:ads_pol_dynamics_bimolecular,pefferkon:ads_pol_exhange_kinetics,%
varoqui:mobility_ads_pol_review,%
frantzgranick:kinetics_adsorption_desorption_prl,%
frantzgranick:ps__cyclohexane_exchange_macro,
schneidergranick:kinetic_traps_exchange,
johnsongranick:exchange_kinetics_pmmma,%
schneider:granick_bimodal,
douglas:kinetics_pol_ads_jpc,%
frantzgranick:pmma_bound_fraction_early,
sogagranick:ir_dichroism_pdms,
sogagranick:orientattion_trains_loops,
fusantore:peo_competitive_adsorption,fusantore:age_relaxation_peo_exchange,%
mubarekyansantore:N_age_peo_exchange,mubarekyansantore:barrier_peo_exchange,
raviv:pol_mat_comments_combo_epje}.
For $\epsilon$ values of several $\kT$ time scales become so long that
the layer build-up becomes essentially an irreversible process leading
to non-equilibrium structure
\cite{schneider:granick_bimodal,douglas:kinetics_pol_ads_jpc}.
One may speculate this is due to cooperative effects mediated by
mutual topological chain constraints which drastically suppress
mobilities near the surface \cite{raviv:pol_mat_comments_combo_epje}.
Indeed, in the scaling theory \cite{gennes:pols_interface_review} for
equilibrium layers, when $\epsilon$ exceeds $\kT$ the smallest loops
in the layer become as small as the monomer size, $a$. Such small
loops are likely to greatly restrict chain motion.

For physical adsorption processes, large sticking energies originate
in hydrogen bonding, dispersion or dipolar forces or attractions
between charged groups. Most metal and silicon-based surfaces are
normally oxidized and many polymer species form strong hydrogen bonds
with the surface oxygen or silanol groups.  Biopolymers such as
proteins and DNA attach tenaciously to many surfaces due to their
large number of charged and polar groups
\cite{brownbotstein:cdna_array,hladybuijs:protein_adsorption}.  Since
hydrogen bonds, for instance, normally have energies of $4\kT$ and
greater \cite{joestenschaad:book:h_bonding} it is apparent that strong
physical bonds are very common.

The extreme example of irreversibility arises in {\em chemi\-sorption}
\cite{lenk:functionalized_pmma_on_gold,schlenoff:functionalized_ps_on_gold,%
shafferchakraborty:pmma_chemisorption_kinetics,adrianichakraborty:kinetic_ising_chemisorption,%
cosgrove:phys_and_chem_adsorption_langmuir} where covalent
surface-polymer bonds develop irreversibly
(fig. \ref{chemiphysi_scheme}).  In various technologies polymers are
attached by chemical reactions to solid surfaces either from solution
as in colloid stabilization by chemically grafting polymers onto
particle surfaces \cite{laiblehamann:chemi_colloid}, or from a polymer
melt as occurs in the reinforcement of polymer-polymer
\cite{creton:kramer:iface_fracture_review,sundararajmacosko:utm,ben:fred}
or polymer-solid
\cite{leewool:adhesion_receptor_density,wu:polymer_iface_adhesion_book,%
edwards:review_filler_reinforcement,creton:kramer:iface_fracture_review}
interfaces.  In general, whether physical or chemical bonding is
involved many applications prefer the strongest and most enduring
interfaces possible and irreversible effects are probably the rule
rather than the exception.

\begin{figure}
\resizebox{\columnwidth}{!}{%
  \includegraphics{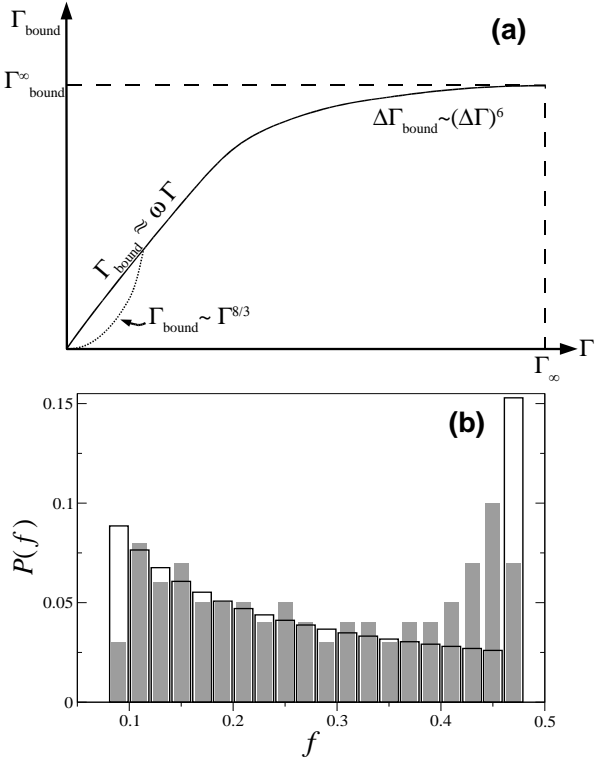}
}
\caption{(a) Theoretically predicted dependence of surface-bound mass,
$\Gammabound$, on total adsorbed mass per unit area, $\Gamma$.  For
irreversible physisorption an initial linear regime crosses over to
sharp saturation $\Delta \Gammabound \twid (\Delta \Gamma)^6$ as
asymptotic coverages are approached.  A similar curve has been
measured experimentally in ref. \cite{schneider:granick_bimodal}.  For
chemisorption the relation is the same, but with an additional initial
regime $\Gammabound\twid \Gamma^{8/3}$.  (b) Probability distribution
of fraction of bound monomers $f$ per chain.  Grey bars reproduce
experimental data from ref. \cite{schneider:granick_bimodal}.  Empty
bars are the theoretically predicted $P(f)\twid f^{-4/5}$ for $f_{\rm
min} < f < \omega$ where values for $f_{\rm min} = 0.09$ and $\omega =
0.47$ were derived from the measurements in
ref. \cite{schneider:granick_bimodal}.  The delta function at $f =
\omega$ is the contribution from early arriving chains and is expected
to be broadened in reality.}
\label{granick_prl}      
\end{figure}

This paper considers dilute solutions. Irreversible adsorption from
melts and semidilute solutions, previously studied both experimentally
\cite{cohenaddad:pdms_melt_silica_coverage_1989,%
cohenaddaddujourdy:pdms_melt_silica_kinetics_1998,auvray:irrev_ads_neutron,%
auvray:irrev_ads_concentrated_solution,ben:pmma_adsorption,marzolin:eng_graft_and_guiseling_brush,%
leger:anchored_chains_adhesion_review,deruelle:pol_modified_adhesion_faraday_discuss}
and
theoretically\cite{guiselin:irrev_ads,cohenaddadgennes:guiselin_brush_poison_elimination,ben:chemiguiselin_letter},
involves rather different processes.  Our aim is to understand the
structure and formation kinetics of layers which are formed under
these irreversible circumstances where the usual statistical
mechanical approach is inapplicable to the non-equilibrium structures
which form.  A series of experiments by Granick and coworkers
\cite{frantzgranick:kinetics_adsorption_desorption_prl,%
frantzgranick:ps__cyclohexane_exchange_macro,
schneidergranick:kinetic_traps_exchange,
schneider:granick_bimodal,douglas:kinetics_pol_ads_jpc,%
johnsongranick:exchange_kinetics_pmmma,%
frantzgranick:pmma_bound_fraction_early,
sogagranick:ir_dichroism_pdms, sogagranick:orientattion_trains_loops}
have examined these issues.  These workers found that when $\epsilon$
reached values of only a few $kT$, polymer relaxation times became
large and equilibrium layers were not attained.  This was most clearly
apparent in experiments following polymethylmethacrylate (PMMA)
adsorption onto oxidized silicon ($\epsilon\approx 4\kT$) via hydrogen
bonding from a dilute CCl$_4$ solution
\cite{johnsongranick:exchange_kinetics_pmmma,frantzgranick:pmma_bound_fraction_early,%
schneider:granick_bimodal,douglas:kinetics_pol_ads_jpc}.  Measuring
both the total adsorbed mass, $\Gamma$, and the surface-bound part,
$\Gammabound$, as a function of time by infrared absorption spectra,
it was found that early arriving chains had much higher fractions of
bound monomers, $f$, than late arrivers, and these $f$ values were
frozen in throughout the experiment's duration of several hours.
Essentially monomers remained irreversibly fixed to the site they
originally landed on, or at least close to this site.  Measuring the
distribution of $f$ values among chains they found a bimodal
distribution shown in fig. \ref{granick_prl}(b) with two peaks at
small and large $f$, respectively.  This is strikingly different to
equilibrium layers where all chains within the layer are statistically
identical and are characterized by the {\em same} time-averaged value,
$f = \Gammaboundinf/\Gammainf$, where $\infty$ denotes asymptotic
values $(t \gt \infty)$.

A number of analytical and numerical efforts
\cite{shafferchakraborty:pmma_chemisorption_kinetics,%
adrianichakraborty:kinetic_ising_chemisorption,%
barford:irrev_ads_1,barfordball:irrev_ads_2,%
konstadinidis:irrev_ads_monte_carlo,%
shaffer:strong_ads_heteropolymers,ponomarev:artem_durning,%
douglas:kinetics_pol_ads_jpc} have
addressed irreversible polymer adsorption from solution.  However, the understanding
of these phenomena remains very far from the quantitative level which
has been achieved for equilibrium layers.  In this paper we develop a
theory which amongst other objectives aims to understand the
experimental findings of refs.
\cite{johnsongranick:exchange_kinetics_pmmma,frantzgranick:pmma_bound_fraction_early,%
schneider:granick_bimodal,douglas:kinetics_pol_ads_jpc}.  We emphasize
adsorption from a dilute polymer solution under good solvent
conditions and consider systems where monomer-surface bonds are strong
enough that they are effectively irreversible within the experimental
timescales. Certain results are generalized to theta solvent solutions.  We
calculate the relationship $\Gammabound(\Gamma)$, how each of
$\Gammabound$ and $\Gamma$ depends on time, as well as the final
layer's distribution of chain contact fractions, $P(f)$.  We will
compare our predicted final layer structure to the well established
theoretical results for equilibrium layers which predict a density
profile $c(z) \twid z^{-4/3}$, and a self-similar loop distribution,
$\Omega(s) \twid s^{-11/5}$.  Similar to the picture that was
developed by Granick and co-workers, we find that the final layer
consists of two populations: early arriving chains lie flat on the
surface while late arrivers can only adsorb with an ever-decreasing
fraction of their monomers onto the surface leading to a diffuse
weakly attached outer layer.  We find a universal distribution of $f$
values, $P(f) \twid f^{-4/5}$ for small $f$ which agrees rather
closely with the experiments of refs.
\cite{schneider:granick_bimodal,douglas:kinetics_pol_ads_jpc}.
Interestingly, we find layer loop distributions and density profiles
obeying the same scaling laws as those of equilibrium layers.  Chain
configurations are very different, however, leading to radically
different physical properties of the layer.

Our picture of irreversible polymer adsorption is in some respects
qualitatively similar to the theoretical one of the workers of
ref. \cite{douglas:kinetics_pol_ads_jpc} who simulated their
experiments in a random sequential adsorption \cite{talbot:rsa_review}
framework.  They visualized chains as ``deformable droplets'': at the
early stages when the coils arrive onto a bare surface, each droplet
adsorbs a certain maximum cross-sectional area.  As available surface
area for adsorption become scarce, in order for late-arriving chains
to adsorb, it was assumed droplets deform by reducing their
cross-sectional areas parallel to the surface to fit into the empty
space. In so doing, they become more extended into the bulk.  Using
this model they generated a $P(f)$ similar to the experimental one of
fig. \ref{granick_prl}(b).  The picture developed here differs in this
respect: late arriving chains freely overhang early flat-lying chains
and rather than fit into a single available connected surface area
they attach at disconnected empty surface sites.  

\begin{figure}
\centering
\resizebox{0.75\columnwidth}{!}{%
  \includegraphics{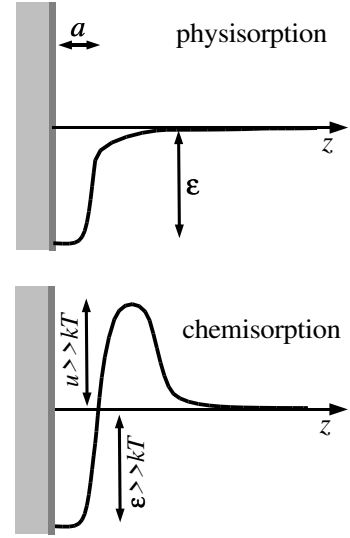}
}
\caption{The two classes studied in this paper: physisorption and
chemisorption.  Simplified view of monomer free energy as a function
of distance between monomer and surface.  In physisorption there is no
activation barrier and monomer-surface association is immediate upon
contact.  When sticking energy $\epsilon$ exceeds a few $kT$,
experiment indicates that effective desorption rates become very
small, presumably due to complex many-chain effects.  Chemisorption
typically involves a large activation barrier, $u \gg kT$, which needs
many monomer-surface collisions to traverse. The adsorption strength
is also large, $\epsilon \gg kT$. Some
systems are in practice mixtures of chemi and physisorption, a
complexity  not dealt with in the present work.}
\label{well_barrier}      
\end{figure}

The process of irreversible polymer layer formation entails
progressive freezing in of constraints due to irreversible
monomer-surface bonding.  These constraints gradually reduce the
volume of configurational phase space available.  Thus ergodicity is
inapplicable and in lieu of the algorithms of equilibrium statistical
mechanics one must follow the kinetics of chain adsorption and layer
build up.  It is important to distinguish carefully between two broad
classes of adsorption kinetics, physisorption and chemisorption, which
are characterized by very different values of the local reaction rate
$Q$. We define this as the conditional monomer-surface reaction rate,
given an unattached monomer contacts the surface (see
fig. \ref{well_barrier}).  In physisorption, monomer attachment is
essentially diffusion-limited, $Q\approx 1/\ta$, where $\ta$ is
monomer relaxation time.  Chemisorption, where adsorbed monomers form
very strong chemical bonds with the surface
\cite{konstandinidis:pmma_chemisorprion,lenk:functionalized_pmma_on_gold,%
schlenoff:functionalized_ps_on_gold,%
shafferchakraborty:pmma_chemisorption_kinetics,%
adrianichakraborty:kinetic_ising_chemisorption,%
cosgrove:phys_and_chem_adsorption_langmuir}
is much slower \cite{ben:fred_letter,ben:fred,ben:grosberg_book} with $Q$ values 8 or
more orders of magnitude smaller than those of physisorption.  The
origin of this difference is that chemical bond formation usually
involves a large activation barrier (fig. \ref{well_barrier}).

Our theoretical
approach is to make the simple assumption of total irreversibility,
motivated by the experiment discussed above: once a monomer bonds with
the surface, our model is that this bond never breaks.  The processes
of chemisorption and physisorption will be analyzed separately.
Though we find both lead to similar final layer structures, the
kinetics are very different.  In physisorption a single chain adsorbs
onto a bare surface in a time of order the coil relaxation time or
less, typically microseconds, whereas single chain chemisorption may
last minutes or even hours and is thus observable experimentally.  We
begin with a study of chemisorption in sections 2, 3, and 4.
Specifically, in Section 2 we show that quite generally there are
three possible modes of single chain adsorption, depending on the
value of a certain critical exponent $\theta$.  Single chain
chemisorption from good and theta solvents is then studied in detail
in Section 3.  In Section 4 we consider the later stages of the
kinetics when chains overlap and dense layers are formed
(fig. \ref{chemiphysi_scheme}).  The case of irreversible
physisorption is treated in Section 5.  We compare our results to
experiment and conclude with a discussion in Section 6.
An announcement of our results has appeared in 
ref. \cite{ben:chemiphysi_letter}.


\section{Chemisorption, General Discussion: Three Modes of Adsorption}

\begin{figure}
\resizebox{\columnwidth}{!}{%
  \includegraphics{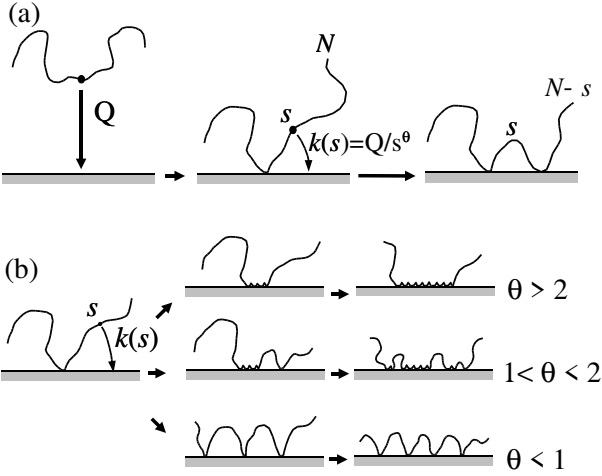}
}
\caption{Schematic of single chain chemisorption.  (a) A chain makes
its first attachment with the surface and starts its adsorption
process.  The rate of first monomer attachment is proportional to $Q$,
namely the reaction rate given the monomer touches the surface.
Subsequent monomer adsorption is determined by the form of the
reaction rate $k(s)$ for the $s\nth$ monomer away from the first
attachment.  For $s \ll N$, where $N$ is length of the tail, $k
\approx Q/s^{\theta}$ where the exponent $\theta$ reflects equilibrium
polymer statistics near interfaces, given the constraints imposed by
earlier reactions.  (b) Depending on $\theta$, three modes of
adsorption are possible. (i) Zipping, $\theta > 2$.  Monomers adsorb
sequentially forming a long train.  (ii) Accelerated zipping,
$1<\theta<2$.  As (i) but zipping is accelerated by nucleation of new
zipping centers due to formation of large loops.  (iii) Uniform
collapse ($\theta < 1$).  Monomers distant from adsorbed ones come
down first leading to formation of large loops whose size is uniformly
decreasing with time.  Chemisorption from dilute solution with good
solvent is accelerated zipping ($\theta=8/5$).
}
\label{single_chain_story}
\end{figure}

Our interest is an initially bare surface confronting a dilute
solution of functionalized chains. During the earlier stages of
surface layer formation, the coverage is small and individual chains
do not see each other. In this section we consider in detail how a
single chain chemisorbs onto an interface, remembering that
chemisorption is characterized by very small values of the local
monomer-surface reaction rate $Q$. The chain will make an initial
attachment and then develop a certain loop profile as successive
monomers gradually attach, eventually leading to complete chain
adsorption in a certain time $\tauads$ (see
fig. \ref{single_chain_story}) whose dependence on chain length is an
important characteristic.  Since the early polymer layer consists of
chains dilute on the surface as in fig. \ref{chemiphysi_scheme}(a),
the initial layer structure is a superposition of these single chain
loop profiles.

In this section we consider chemisorption in its full generality.  We
show that there are three distinct classes of behavior, each with
different adsorption modes, loop profiles and adsorption times
$\tauads$. Which class a given experimental system belongs to depends
on the bulk concentration regime and other factors. The classes are
defined by the value of a critical exponent characterizing polymer
statistics near surfaces.

We will assume that all monomers are identical and chemically active,
and that the surface has a free energetic preference for the solvent
over the polymer species. This means that in terms of physical
interactions the polymers see a hard, repulsive wall.  We choose units
where the monomer size $a$ is unity.

Consider the chain in fig. \ref{single_chain_story}(b) whose first
monomer has just reacted with the surface.  Which of the chain's
remaining monomers will react next?  What is the sequence of
chemisorption events? This depends on the form of the reaction rate
$k(s)$ at which a reactive group $s$ monomer units along the backbone
away from the bound monomer reacts with the surface, as shown in
fig. \ref{single_chain_story}(a).  Now the important feature of
chemisorption is that due to the smallness of $Q$ in order for any
chemical bond to form a time much larger than the attached chain's
relaxation time is needed.  Thus quite generally the reaction rate for
a given monomer at any moment is proportional to the equilibrium
contact probability of finding this monomer on the surface given the
current constraints due to all chemical bonds formed at earlier
times. In this particular situation, this is
                                                \begin{eq}{first}
p(s|N) =   {Z(s,N) \over Z(N)} \comma
                                                                \end{eq}
where $Z(N)$ is the partition function of the chain with one monomer
bound (middle of fig. \ref{single_chain_story}(a)) and $Z(s,N)$ is the
partition function of the chain which has the additional constraint
that the $s\nth$ monomer is also bound (last of
fig. \ref{single_chain_story}(a)).  Physically one expects that for
small enough $s$, this is independent of $N$, \ie $p(s|N) \approx
s^{-\theta}$ where the value of the exponent $\theta$ reflects the
equilibrium polymer statistics, given the constraints.  This then
leads to the following expression for the reaction rate:
                                                \begin{eq}{kse}
k(s) \approx  {Q \over s^\theta} \gap (s \ll N) \period
                                                                \end{eq}
The total reaction rate for the next adsorption event is a sum over
all the $N-1$ monomers which may adsorb next. These belong to the two
tails in fig. \ref{single_chain_story}(b) which are of order $N$ in
length.  The net reaction rate is thus approximately
                                                \begin{eq}{nice}
\Rtotal \approx \int_1^N k(s) ds 
                                                                \end{eq}
which, depending on the value of $\theta$, is dominated by either
small or large $s$.

This argument is then repeated for the reactions of the second, third
and subsequent monomers, in all cases described by a rate with the
same small-$s$ behavior as in eq. \eqref{kse}.  The only difference is
that tails are replaced by loops.  Thus the entire kinetic sequence is
characterized by the single exponent $\theta$.  The nature of the
kinetics depends on the value of $\theta$ as follows.

(i) $\theta > 2$. Suppose the equilibrium statistics
(eq. \eqref{first}) are such that $k(s)$ decays faster than $1/s^2$.
Then $\Rtotal \approx k(1)$ and typically a monomer near the first
attachment is most likely to attach next.  The third monomer to
attach, repeating the same argument, will be near the first two,
and so on.  In this case therefore the chain would {\em zip} onto the
surface starting at the first attachment point, as shown in the top of
fig. \ref{single_chain_story}(b).  Since each new attachment occurs at
the same rate each time, the full zipping time would then be $\tauads
\approx N / k(1) \approx Q^{-1} N$.

(ii) $\theta < 1$.  In this case $\Rtotal$ is dominated by $s$ of
order $N$ reflecting the fact that even though a monomer with small
$s$ is on average more frequently near the surface than a distant one,
there are many more distant ones and their number is the dominating
factor.  Thus a distant monomer of order $N$ units away from the first
graft attaches next, leading to the formation of a loop of size of
order $N$.  Repeating the argument for the subsequent reactions leads
to a process of uniform {\em collapse} of the chain where roughly
centrally positioned monomers of loops react with the surface at each
step, as shown in the bottom case of fig. \ref{single_chain_story}.
In this case the average lifetime of a loop with $s$ monomers is
$\tau_{\rm loop}(s) \approx 1/\int_1^s ds' k(s') \approx Q^{-1}
s^{1-\theta}$.  Thus smaller loops take longer time to collapse and
the rate limiting step for full adsorption is the collapse of the
smallest loops: $\tauads \approx \tau_{\rm loop}(1) \approx Q^{-1}$.

(iii) $1 < \theta < 2$. This is intermediate between zipping and
collapse.  Though $\Rtotal$ is dominated by small $s$, suggesting
zipping, there is in fact enough time before pure zipping completes
for large loops to form.  To see this, return to the 1-graft situation
on the extreme left of fig. \ref{single_chain_story}(b). The time for
a loop of order $N$ (for argument's sake, bigger than $N/2$) to form
from one of the two tails extending from the grafted monomer is roughly:
                                                \begin{eq}{brown}
\tauN \approx \paren{
\int_{N/2}^N ds\, k(s) 
}^{-1}  \approx Q^{-1} N^{\theta-1} \period
                                                                \end{eq}
This is much smaller than $\tauads \twid N$ which pure zipping would
give, and thus before puring zipping is complete, large loops must
have formed.  We call this case {\em accelerated zipping} since large
loop formation effectively short-circuits the pure zipping process by
nucleating new sources of zipping as shown in the middle case of
fig. \ref{single_chain_story}(b). Now eq. \eqref{brown} implies
(unlike $\theta<1$) larger loops take longer to form.  Thus by $\tauN$
loops of all sizes have come down, \ie the chain has adsorbed and we
conclude $\tauads \approx \tauN$.

In summary, depending on the value of $\theta$ the chemisorption time
has three possible forms:
                                                \begin{eq}{tauads-summary}
Q \tauads \approx \casesbracketsshortiii
{{\rm const.}} {\theta < 1}
{N^{\theta-1}}  {1 < \theta < 2}
{N}  {\theta > 2}
                                                                \end{eq}
In the next section we calculate the value of $\theta$ under good
solvent conditions and find that it belongs to the accelerated
zipping case.  We then study the corresponding case in detail.


\section{Single Chain Chemisorption in a Good Solvent: Accelerated Zipping}

In this section we consider the kinetics of single-chain
chemisorption, describing polymer layers during the early
stages of the chemisorption process, fig. \ref{chemiphysi_scheme}(a).
Expressions are derived for the number of bound monomers,
$\gammabound(\tau)$, and the single-chain loop distribution profile,
$\omega_\tau(s)$, where $\tau$ measures time after first monomer
attachment onto the surface.  In order to perform these calculations, in
this section we first evaluate the reaction exponent $\theta$ of
eqs. \eqref{kse} and \eqref{tauads-summary} and find it belongs to the
accelerated zipping class.  Subsequently we solve the accelerated
zipping kinetics, first using simple scaling arguments and then a
more detailed solution of the rate equations.

\subsection{The Reaction Exponent $\theta$}
        
\begin{figure}
\centering
\resizebox{0.85\columnwidth}{!}{%
  \includegraphics{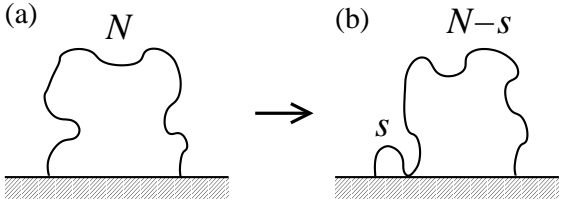}
}
\caption{A polymer loop anchored to the surface by its two ends.  The
reaction rate of its $s\nth$ monomer with the surface, leading to two
loops of length $s$ and $N-s$, is proportional to the ratio of the
partition functions of the loop before and after the reaction. 
}
\label{mother_daughter}      
\end{figure}

To our knowledge $\theta$ in good solvents has not been calculated
before.  We show here that it can be expressed in terms of other
known critical exponents characterizing polymer networks
\cite{duplantier:networks,debelllookman:surface_exponents_review}.
The simplest way to derive $\theta$ is to consider a loop of $N$
monomers bound to the surface by its two ends as in
fig. \ref{mother_daughter}(a).  Then the reaction rate of its $s\nth$
monomer resulting in the formation of two loops of lengths $s$ and
$N-s$, as shown in fig. \ref{mother_daughter}(b), is
                                                \begin{eq}{kfull}
k(s|N)  =  Q {\Zloop(s,N-s) \over  \Zloop(N)}  
                                                                \end{eq}
where $\Zloop(N)$ and $\Zloop(s,N-s)$ are the partition functions of
the loops of fig. \ref{mother_daughter}(a) and
\ref{mother_daughter}(b), respectively.  
Now although $\Zloop(N)$ is known, $\Zloop(s,N)$ is
only known for $s$ of order $N$.  From ref. \cite{duplantier:networks}:
                                                \begin{eq}{bad-day}
\Zloop(N) \approx \mu^N N^{\gamma_a -1} 
\comma \ \ 
\Zloop(N/2,N/2) \approx \mu^N N^{\gamma_c -1} \comma
                                                                \end{eq}
where $\mu$ is a
constant of order unity and the values of of $\gamma_a,\gamma_c$ have
been calculated analytically and numerically
\cite{duplantier:networks,debelllookman:surface_exponents_review}.
Now demanding that $\Zloop(s,N-s)$  follows a power law
for small $s$ and that it reduces to eqs. \eqref{bad-day} in the $s \gt
1$ and $s \gt N/2$ limits one has 
                                                \begin{eq}{scaling-assumption}
\Zloop(s,N-s) \approx \mu^{N} N^{\gamma_c-1} \zeta\paren{s \over N} 
\period
                                                    \end{eq}
Here the scaling function $\zeta$ must obey:
                                                \begin{eq}{zeta}
\zeta(x) \approx \casesbracketsshortiii{x^{-\theta}}  {x \gt 0}
                                        {1}            {x = 1/2}
                                        {(1-x)^{-\theta}} {x \gt 1}
                                                                \end{eq}
where the $x \gt 0$ and $x \gt 1$ behaviors are identical by symmetry and
                                                \begin{eq}{theta}
\theta = 1 + \nu \approx 8/5 \ \ \ \ {\rm (good\ solvent)}
                                                                \end{eq}
which follows after using the identity \cite{duplantier:networks}
$\gamma_a-\gamma_c = \nu +1$.  Here $\nu \approx 3/5$ is the
Flory exponent \cite{gennes:book}.  From eqs. \eqref{kfull},
\eqref{bad-day}, and \eqref{scaling-assumption} one thus has
                                                \begin{eq}{superdil}
k(s|N) \ \approx \ {Q \over N^\theta} \ \zeta \paren{s \over N}
\comma
                                                                \end{eq}
which for $s \ll N$, reduces to eq. \eqref{kse}, $k \twid
s^{-\theta}$.  Thus single chain chemisorption in a good solvent is
characterized by $\theta \approx 8/5$; since $1 < \theta < 2$ the
adsorption mode is accelerated zipping.

Eq. \eqref{superdil} was derived for the particular case of a loop as
in fig. \ref{mother_daughter}.  But even if the reacting monomer is
part of a tail as in fig. \ref{single_chain_story}(a), by a
straightforward generalization of the reasoning of this section using
the exponents calculated in ref. \cite{duplantier:networks}, one can
show that the reaction rate has the same form for $s \ll N$.  One can
show the same is true if the ends of the loop in
fig. \ref{mother_daughter} are connected to other loops.

The above results for $\theta$ directly generalize to theta solvent
solutions.  In this case polymer statistics are effectively ideal
\cite{gennes:book} and $\nu \gt 1/2$ in all of the above, \ie
                                                \begin{eq}{theta-theta}
\theta = 3/2  \gap ({\rm theta\ solvent})
                                                                \end{eq}
which also corresponds to accelerated zipping.  In fact, for this case
the exponent can be obtained more simply as follows: the probability
that the $s\nth$ monomer, measured from a given surface attachment, is
in contact with a hard wall is proportional to the return probability
$P$ of a random walk which starts one step away from an absorbing
surface after $s$ steps, \ie $k \twid P \twid s^{-3/2}$.

Finally we remark that the partition functions of eq. \eqref{bad-day}
correspond to loops such as those of fig. \ref{mother_daughter} whose
end locations on the surface are annealed \cite{duplantier:networks}.
However for chemisorption onto solid surfaces the ends are either
completely fixed or may diffuse very slowly on the surface.  (Note
that chemisorption onto liquid interfaces are cases where the end
locations are truly annealed.)  Nonetheless, all our results for $s
\ll N$ and the scaling structure of eq. \eqref{superdil} must remain
the same even in this case since in the $s \ll N$ limit, $k(s|N)$ is
independent of the location of the other end.  Our general conclusions
are thus expected to be robust, provided the two ends are not so far
apart that there are strong lateral loop stretching effects.  In the
following we self-consistently assume that as the adsorption process
proceeds there is no tendency for generated loops to be in such
stretched configurations.

\subsection{Accelerated Zipping: Scaling Analysis}

Let us consider now the kinetics of adsorption starting from a polymer
chain as in fig. \ref{single_chain_story}(a) which has just made its
first attachment with the surface with an interior monomer.  This is
in fact typical: we show in Appendix A that chains are much more
likely (by a factor $N^{0.48}$) to make their first
surface contact with an interior monomer because there are many more
(of order $N$) such monomers as compared to chain ends.

Now the chain starts to chemisorb at $\tau=0$.  From the previous
subsection we have seen that in the accelerated zipping mode the chain
adsorption time $\tauads$ is equal to $\tauN$, the timescale
associated with loops of size $N$.  What will the chain's loop profile
be for times $\tau$ smaller than $\tau_N$?  Initially, since the
reaction rate is dominated by small $s$, the chain will start to zip
and sequences of bound monomers (``trains'') will grow outwards from
the first attachment point.  As time proceeds though, large loops
start to come down due to adsorption of monomers distant to the
already bound ones.  These loops have a certain distribution of loop
sizes $\omegatau(s)$ and their effect is to accelerate the zipping
process since they nucleate further sources of zipping and train
formation.  The maximum size of such loops which had enough time to
form by time $\tau$, \ie for which $\int_{\staumax}^N ds' k(s')
\approx 1/\tau$, is
                                                \begin{eq}{ss}
\staumax \approx (Q \, \tau)^{1/\nu}
\period
                                                                \end{eq}
which increases with time.  The maximum loop size thus becomes of
order $N$ at $\tauN$.

\begin{figure}
\centering
\resizebox{0.9\columnwidth}{!}{%
  \includegraphics{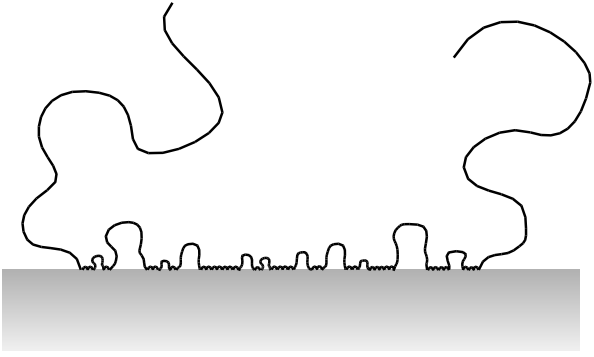}
}
\caption{Snapshot of loop profile of a chemisorbing chain during its
accelerated zipping down onto a surface.  At an earlier time the chain
adsorbed with an interior monomer from which zipping started to
propagate outwards accompanied by occasional formation of large loops
which nucleated further zipping sources as shown.  The distribution of
loop sizes follows a universal power law and as larger loops form
with increasing time, loop sizes are larger near the outermost tails.
The tails act as a continuous loop source thus sustaining the loop
distribution.  Eventually, after $\tauads \approx Q^{-1} N^{\nu}$ the
largest loop becomes of order the tail size, $N$, the tails are
completely consumed and the structure collapses onto a completely flat
configuration.}
\label{acc_zipping}      
\end{figure}

Overall, we expect the chain to consist of three parts, shown in
fig. \ref{acc_zipping}: (i) Two large tails, each of length of order
$N$.  These are the two tails of the polymer chain which was initially
bound by one of its interior monomers as in
fig. \ref{single_chain_story}(a).  For $t \ll \tau_N$ the tails had
neither enough time to decay into large loops, nor to completely
zip. (ii) Loops with a loop distribution $\omegatau(s)$. (iii) Trains
of bound monomers whose number we define to be $\gammabound(\tau)$.

Now $\gammabound(\tau)$ is easily found by making the ansatz that it
follows a power law in time.  In addition, for short times
$\gammabound$ is independent of $N$; one can imagine sending the chain
size to infinity, which would not affect the accelerated zipping
propagating outwards from the initial graft point.  This dictates:
                                                \begin{eq}{mtrains}
\gammabound(\tau) \approx N \paren{\tau \over \tau_N}^{1/\nu}
\gap
(\tau \ll \tau_N)
                                                                \end{eq}
since by $\tau_N$ most of the chain has adsorbed, \ie we must have 
$\gammabound (\tau_N) \approx N$.

We now evaluate $\omegatau(s)$ by making the ansatz that it also has
 power law behavior:
                                                \begin{eq}{delta-loop}
\omegatau (s)
\ \approx \
\inverse{\staumax} \paren{\staumax \over s}^\delta
\gap
(1 \ll s \ll \staumax) 
                                                                \end{eq}
with $\delta > 1$.  Since by $\tau$ there has been enough time for the
formation of order one loop of length $\staumax$, the normalization
has been chosen such that there is one loop of order $\staumax$ in size.
Now the total number of loops is thus $L(\tau) = \int_1^N ds\,
\omegatau(s) \approx (\staumax)^{\delta-1} $.  These loops provide
$L(\tau)$ new nucleation points for further zipping.  The reaction
rate at each such point is dominated by small $s$ (since $\theta > 1$)
and is thus of order $k(1)$, so
                                                \begin{eq}{pants}
{d \gammabound\over d\tau} \approx  k(1) L(\tau) 
\twid \tau^{(\delta -1)/\nu } 
\gap (\tau \ll \tau_N) \period
                                                                \end{eq}
Now demanding that \eqref{pants} reproduces eq. \eqref{mtrains} the
value of $\delta$ is determined:
                                                \begin{eq}{delta}
\delta = 2 - \nu
\period
                                                                \end{eq}
Notice that the above results are self-consistent: the number of loops
$L(\tau)$ is much smaller than $\gammabound(\tau)$, and essentially
all reacted monomers do belong to trains as assumed in
eq. \eqref{mtrains}.

Let us summarize this subsection's results, setting $\nu =3/5$ which
is the value relevant to our main interest of dilute solutions with
good solvent. From eqs. \eqref{ss}, \eqref{mtrains} and
\eqref{delta-loop} we have:
                                                \begin{eq}{summary-scaling}
\gammabound(\tau) \approx (Q \tau)^{5/3} \comma \ \ \ 
\omegatau(s) \approx (Q \tau)^{2/3} s^{-7/5}
                                                                \end{eq}
valid for $\tau < \tauN$ and $s < \staumax  \approx (Q \tau)^{5/3}$.

\subsection{Accelerated Zipping: Solution of Rate Equations}

In this section we analyze the accelerated zipping phenomenon in a
more rigorous way by solving the loop evolution dynamics.  We will
recover all of the scaling results of section 3.2.  The time evolution
of the chain's loop distribution is given by
\cite{shafferchakraborty:pmma_chemisorption_kinetics}
                                                \begin{eq}{loops}
{d\omegatau(s)\over d \tau} = 2 \int_{s}^N ds'\, \omegatau(s')\, k(s|s') 
   - \int_0^s ds'\, \omegatau(s)\, k(s'|s) 
\comma
                                                                \end{eq}
where the first term on the right hand side (rhs) describes the rate
of formation of loops of length $s$ by bigger ones, while the second
term on the rhs is the rate of decay of a loop of length $s$ into
smaller loops \cite{chemiphysi_euro:note_fragmentation} and $k$ is
given by eq. \eqref{superdil}.  The initial condition for
eq. \eqref{loops} is $\omega_0(s) = \delta (s-N)$, \ie there is only
one initial loop of length $N$ (in the following for simplicity we do
not distinguish between tails and loops).  A basic assumption in
eq. \eqref{loops} is that $k(s|N)$ retains the form of
eq. \eqref{superdil} throughout the adsorption process. That is, we
assume topological and many-loop excluded volume effects (beyond those
contained in eq. \eqref{kfull}) are not strong enough to alter the
essential form of $k(s|N)$.  It is easy to show by integrating
eq. \eqref{loops} over all $s$ (see calculation in Appendix B) that
the kinetics conserve the total number of monomers.
Formally, $\omegatau(s)$ in eq. \eqref{loops} is an ensemble average
over many chains.  However the number of loops formed by a single
chain becomes much larger than unity as time increases and thus
eq. \eqref{loops} accurately describes the loop distribution of a
single chain as well.

Although both integrals in eq. \eqref{loops} diverge at $s'=s$ and
$s'=0$, these divergences cancel with one another.  In fact defining
$\Dtau(s)$ to be the number of loops larger than $s$,
                                                \begin{eq}{d}
\Dtau(s) \ \equiv \ \int_s^{\infty} ds' \, \omegatau(s')
                                                                \end{eq}
we obtain the following equivalent rate equation which has
well-defined integrals:
                                                \begin{eq}{deq}
{d \Dtau(s) \over d\tau} 
\ = \
\int_s^N ds' \, \lambda(s|s') \, \omegatau(s') \comma
                                                                \end{eq}
where
                                                \begin{eqarray}{lambda}
\lambda(s|s') &\equiv& {Q \over s'^{\nu}}
\int_{s/s'}^{1-s/s'} dx \, \zeta(x)
                                                        \ddrop
\ &=& \ 
\casesbracketsshortii{- A Q/(s'-s)^{\nu}\ \ \ }         {s' \gt s}
                     {B Q/s^{\nu}}   {s' \gg s}
                                                                \end{eqarray}
with $A,B$ positive constants of order unity.  Eq. \eqref{deq} is
derived by integration of eq. \eqref{loops}.  Its validity is easy to
check by differentiation with respect to $s$ and using $\zeta(s/s') =
\zeta(1-s/s')$; one then recovers eq. \eqref{loops}.  Notice that the
function $\lambda(s|s')$, which describes how much each loop $s'$
contributes to changes in $\Dtau(s)$, is negative for $s'<2s$ since a
loop shorter than $2s$ always creates at least one loop smaller than
$s$.  Also $\lambda(s|s')$ is positive for $s' > 2s$ since at least
one loop bigger than $s$ is created by a loop longer than $2s$.

\begin{figure}
\resizebox{\columnwidth}{!}{%
  \includegraphics{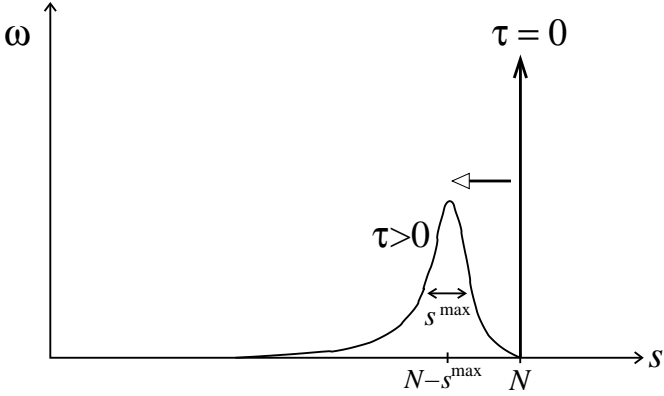}
}
\caption{ Sketch of the size probability distribution $\omegatau(s)$
for the large tail, ($s \gt N$, fig \ref{acc_zipping}) for a polymer
chemisorbing from a good solvent, see eq. \eqref{mother}.  The initial
($\tau=0$) tail length is a $\delta$-function peaked at $N$ and
decreases with time $\tau$ on average by $\staumax$.  The broad
distribution at $\tau$ represents fluctuations in tail length among a
population of many chains.  The normalization of the distribution
remains of order unity for $\tau \ll \tauN$.}
\label{omega_big}      
\end{figure}

We saw in subsection 3.2 that the two large tails forming after the
first attachment have of order $N$ monomers until essentially the
end of the chemisorption process.  In order to describe the tail
kinetics, let us consider $s$ very close to $N$ in eq. \eqref{deq}:
                                                \begin{eq}{rum}
{d \Dtau(s) \over d\tau}
       \approx
- \int_s^N ds' {A Q \over (s'-s)^{\nu}} \omega_\tau(s')
\ \ \ \ \
(s \gt N) \period
                                                                \end{eq}
Self-consistently, one can show that error terms arising from
approximating $\lambda$ by $- A Q /(s'-s)^{\nu}$ in eq. \eqref{rum}
are higher order. Eq. \eqref{rum} is solved in Appendix C, with
solution
                                                \begin{eqarray}{mother}
\omegatau(s)  &=&
\inverse{\staumax} \paren{\staumax \over N-s}^{1 + \nu}
h \paren{\staumax \over N-s} \comma\ \   (N-s \ll N)
                                        \ddrop
{\rm where} && \gap h(x) \gt \casesbracketsshortii{0}               {x \gg 1}
                              {\rm const.} {x \ll 1}    
                                                                \end{eqarray}
The cutoff function $h(x)$ decays to zero exponentially rapidly as $x
\gt \infty$.  Eq. \eqref{mother} (see fig. \ref{omega_big}) describes
the length of the two initial tails, which is continuously decreasing.
After time $\tau$ the length has decreased by $\staumax$.  This
decrease is the maximum loop size that any tail could have created by
time $\tau$.  Due to fluctuations, the initial $\delta$-function peak
broadens as it moves towards $s=0$.  Thus $\omegatau(s)$ in
eq. \eqref{mother} is the probability distribution of the tail length.
One sees that by time $\tauN$ the length of the tail (the largest
non-adsorbed chain segment) shrinks to zero, thus marking the end of
the single chain adsorption process.

Consider now loops ($1 \ll s \ll N$) generated by occasional grafting
of distant monomers.  For short enough times the only sources of loop
formation are the tails which dominate the integral on the rhs of
eq. \eqref{deq} at $s'$ of order $N$.  Assuming for the moment that
this continues to be true for all times, one has from eq. \eqref{deq}
                                                \begin{eq}{big-one}
{d \Dtau (s) \over d\tau}
\ \approx \ 
{Q \over s^{\nu}}
\comma \gap
(\staumax \ll s \ll N)
                                                                \end{eq}
which after integration gives
                                                \begin{eq}{bad}
\omegatau(s) \approx {1 \over \staumax} \paren{\staumax \over s}^{1 + \nu}
\ \ \ \ \ (\staumax \ll s \ll N)
\period
                                                                \end{eq}
Thus the first loops which form have the same loop distribution
profile as the small $s$ behavior of $k(s|N)$.  However, the validity
of eq. \eqref{bad} is limited to $s > \staumax$.  This can be seen by
substituting back eq. \eqref{bad} into eq. \eqref{deq}, revealing that
the assumption that only tails contribute to the rhs of
eq. \eqref{deq} is incorrect for $s \lsim \staumax$.  This limits the
validity of eq. \eqref{bad} to those loop sizes too great to have been
formed by tail collapse events by the time $\tau$.  This is reflected
by the fact that the integral of eq. \eqref{bad} in its region of
validity is of order unity and dominated by $s$ of order $\staumax$.

\begin{figure}
\resizebox{\columnwidth}{!}{%
  \includegraphics{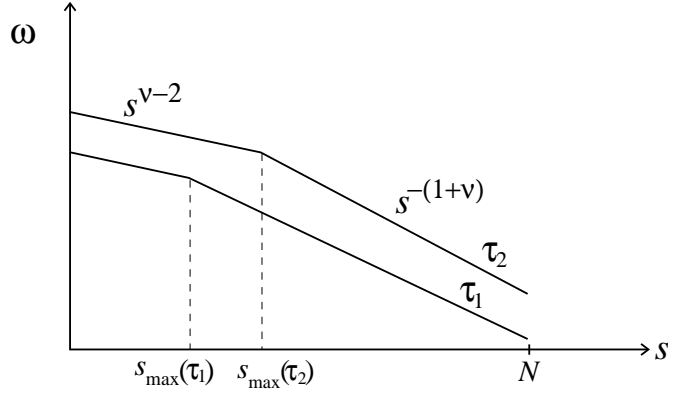}
}
\caption{Log-Log plot of single-chain loop distribution $\omegatau(s)$
for chemisorption from dilute solutions at two different times,
$\tau_1 < \tau_2$.  The distribution consists of an inner and an outer
region with $\omega \twid s^{-7/5}$ and $\omega \twid s^{-8/5}$,
respectively, for good solvents ($\nu=3/5$).  Both the normalization
which is equal to the total number of loops, and the size of the inner
region increase with time.  The outer region extends up to $s$ of
order $N - \smax$.  For larger $s$ the distribution is shown in
fig. \ref{omega_big} describing the size of the large tail.}
\label{omega_loops}      
\end{figure}

Now for $s < \staumax$, the main body of the loop distribution, we
seek a quasi-static solution, $d \Dtau /d \tau \approx 0$, or equivalently
from eq. \eqref{deq},
                                                \begin{eq}{small-one}
\int_s^N ds' \, \lambda(s|s') \, \omegatau(s') \approx 0
\gap
{(1 \ll s \ll \staumax)} 
                                                                \end{eq}
representing the expectation that such loops reach a self-maintained
universal distribution.  We show in Appendix D that
eq. \eqref{small-one} has a power law solution:
                                                \begin{eq}{good}
\omegatau(s) \approx {1 \over \staumax} \paren{\staumax \over s}^{2- \nu}
\gap (1 \ll s \ll \staumax) \period
                                                                \end{eq}
The normalization of eq. \eqref{good} was fixed by demanding
continuity with eq. \eqref{bad} at $\staumax$.  Combining
eq. \eqref{good} and eq. \eqref{bad} we obtain the overall loop
distribution shown in fig. \ref{omega_loops}.

Now the remaining part of the loop profile is the particular
contribution due to trains. That is, we must still calculate the total
number of adsorbed chain units $\gammabound(\tau)$. One might think
this is just $\omega(1)$, the number of minimal length loops. This is
wrong because there is in fact a sink for loops of length 1, \ie a
current out of the $\omega(s)$ distribution into the total mass of
trains, $\gammabound(\tau)$. The latter obeys the dynamics of
eq. \eqref{loops} but with the decay term deleted since bound monomers
cannot decay into smaller ones:
                                                \begin{eq}{ball}
{d \gammabound \over d\tau} \approx 
2 \int_{1}^{\infty} ds'\, \omegatau(s')\, k(s|s') 
\approx Q^{-1} \, L(\tau) \period
                                                                \end{eq}
Here unlike eq. \eqref{loops} we used an explicit cutoff at $s=1$.
In the last expression of eq. \eqref{ball} we used the definition of
the total number of loops, $L(\tau)$, and took into account that the
integral is dominated by its lower limit.  Thus eq. \eqref{ball}
reduces to eq. \eqref{pants} leading to the solution of
eq. \eqref{mtrains}.

In summary, using $\nu = 3/5$, our results for single chain
chemisorption are
                                                \begin{eqarray}{summary-dil}
&& \gammabound(\tau) \approx 
\casesbracketsshortii {(Q \tau)^{5/3}}  {\tau < \tau_N}
                 {N}             {\tau > \tau_N}
\ \ \ \  
                                                \ddrop
&&  \omegatau(s) \staumax               
  \approx                               \ddrop
&& \ \ \left\{
\begin{array}{llll}
{(\staumax/s)^{7/5}}
                       \      &            {(1 <  s < \staumax)}  \blank
{(\staumax/s)^{8/5}} 
                       \      &             {(\staumax < s < N)}  \blank
{[\staumax/(N-s)]^{8/5} \, h({\staumax/(N-s)}) \  } 
                        \     &         {(s \gt N)}   \blank
\end{array}%
\right.
\drop
                                                                \end{eqarray}
where
                                                \begin{eq}{smax-tau}
\tau_N \approx Q^{-1} N^{3/5} \comma \gap
\staumax \approx (Q \tau)^{5/3}        \comma
                                                                \end{eq}
and the expression for $\omegatau$ is for $\tau < \tauN$.  A schematic
of the loop distribution is shown in figs. \ref{acc_zipping},
\ref{omega_big}, and \ref{omega_loops}.  Apart from extra details
these results are identical to the ones we obtained in section 3.2,
eq. \eqref{summary-scaling}, using scaling arguments.  Analogous
results to eq. \eqref{summary-dil} can be obtained for theta solvents
replacing $\nu \gt 1/2$ in all expressions of this section.


\section{Chemisorption: Kinetics of Polymer Layer Formation}

In this section we consider the full chemisorption process where many
chains simultaneously attach to the surface as in
fig. \ref{chemiphysi_scheme}.  Initially, this is a simple
superposition of single chain adsorption processes discussed in
section 3. As chains build up and overlap, this becomes a many-chain
phenomenon.  We will determine the kinetics of total and surface-bound
mass per unit area, $\Gamma(t)$, and $\Gammabound(t)$, the
relationship $\Gammabound(\Gamma)$, the monomer density profile
$c(z)$, and the internal structure of the layer characterized by the
loop distribution per site, $\Omega(s)$.  Finally we determine the
distribution of fraction of adsorbed monomers per chain, $P(f)$.  We
explicitly set $\nu=3/5$ and it will be convenient to reinstate
explicit reference to the monomer size $a$, hitherto set to unity.

\subsection{Monolayer Forms at Early Stages}

In the early stages of adsorption, surface-attached chains are dilute
on the surface and each one performs its accelerated zipping down in a
time $\tauN$.  Thus the layer is a superposition of such chains which
arrived on the surface at different times, the structure of each being
given by eq. \eqref{summary-dil}.  Now the rate of chain arrival onto
the surface is $a^2 d \Gamma/dt = Q N \phisurf$, where $\Gamma$ is
monomers per unit area and $\phisurf$ is the equilibrium monomer
density at the surface:
                                                \begin{eq}{phisurf}
\phisurf = \phibulk {\Zsurf(N) \over \Zbulk(N)}  = {\phibulk \over N}
\period 
                                                                \end{eq}
Here $\Zsurf$, $\Zbulk$ are the single chain partition functions given
a monomer on the surface or in the bulk, respectively.  Their ratio
has been shown to be $1/N$ in ref. \cite{duplantier:networks}.
Solving for the kinetics one thus obtains
                                                \begin{eq}{key}
\Gamma(t) a^2 = \phibulk Q t \comma\ \ \ \ 
 t < \tsatchem \equiv 1/(Q \phibulk)
                                                                \end{eq}
where $\tsatchem$ is the timescale at which $\Gamma \approx 1/a^2$,
\ie when surface density is of order one monomer per ``site'' of area
$a^2$.  Clearly, as $\tsatchem$ is approached, the kinetics of
eq. \eqref{key} should be drastically modified to take into account the depletion
of available landing sites for new chains.

Notice that $\Gamma(\tauN)/N \approx(\phibulk/\phistar)/ \RF^2 $ where
$\phistar = N^{-4/5}$ is the chain overlap threshold concentration
\cite{gennes:book}.  Thus for dilute solutions, $\phibulk<\phistar$,
attached chains are dilute on the surface at time $\tauN$, the
necessary time for a chain to fully adsorb.  At any moment, therefore,
those chains actually in the process of chemisorbing are dilute on the
surface and so do not interfere with one anothers' adsorption process.
Chains adsorb {\em independently}. (Note that for higher
concentrations, \ie for $\phibulk$ in the semidilute regime, chain
interference effects would be important and the resulting layers
different.)

Although a given adsorbing chain does not see {\em simultaneously}
adosrbing chains, in the later stages of layer formation it will be
affected by those which have {\em previously} adsorbed, because these
diminish the available empty surface sites. This effect becomes strong
after time $\tsatchem$. Some interference of this type will commence
at an earlier stage, when flattened down adsorbed chains first start
to overlap on the surface.  This happens at coverage $\Gamma = N/\RF^2
\approx N^{-1/5} a^2$, which occurs after time $\toverlap \approx
\tsatchem N^{-1/5}$.  This is initially a very small effect since at
time $\toverlap$ the fraction of unavailable sites is small for large
$N$, of order $N^{-1/5}$, so the essential features of the adsorption
kinetics will remain unmodified.  This becomes an order unity effect
only at $\tsatchem$.  Between the two times there is a continuously
increasing interference.  However, for practical values of $N$ this is
such a weak power that the timescales $\toverlap$ and $\tsatchem$ will
be difficult to distinguish.  (In fact for theta-solvents, $\nu \gt
1/2$, the two timescales become the same).

Now the bound component of the attached chains of eq. \eqref{key} is a
sum over the bound mass of each chain which depends on the time it
arrived on the surface: $\Gammabound(t)$ = $\int_0^t dt' \Gamma(t')
\gammabound(t-t') /N $.  Using eqs. \eqref{key} and
\eqref{summary-dil} one thus has
                                                \begin{eqarray}{not-good}
&& \Gammabound(t) a^2 \approx \drop
&&\casesbracketsshortii
        {\phibulk Q N^{3/5} (t/\tauN)^{8/3}}
                                        {t < \tauN}
        {\phibulk Q t}
                                        {\tauN < t < \tsatchem} 
                                                                \end{eqarray}

The short-time kinetics of eqs. \eqref{key} and \eqref{not-good} are
sche\-matically shown in fig. \ref{gamma_kinetics}.  One sees two
regimes in the short time behavior of $\Gammabound$ which are also
reflected in two regimes of $\Gammabound(\Gamma)$.  From
eqs. \eqref{key} and \eqref{not-good} one has
                                                \begin{eq}{bad-good}
\Gammabound =
\left\{
\begin{array}{llll}
{\Gamma(\tauN) [\Gamma/\Gamma(\tauN)]^{8/3},}
         \      &        {\Gamma < \Gamma(\tauN)}
                        \blank
{\omega \Gamma,}
                       \      &   {\Gamma(\tauN) < \Gamma < a^{-2}  }
                          \blank
\end{array}%
\right.
                                                                \end{eq}
This early portion of the $\Gammabound(\Gamma)$ curve is illustrated
in fig. \ref{granick_prl}(a).  In eq. \eqref{bad-good} we explicitly
introduced the proportionality prefactor $\omega$.  This is a
small-scale species-\-dependent constant representing the fact that even
for isolated chains, steric constraints at the monomer level prevent
every mono-mer from bounding.  Thus $\omega$ is the fraction of
monomers which are allowed by such constraints to bound.

In summary, during the early stages of chemisorption, chains flatten
out on the surface uninhibited by the presence of others.  For very
short times, $t < \tauN$, none of the attached chains has completed
its adsorption and the layer's loop distribution is a superposition of
single-chain loop structures given by eq. \eqref{summary-dil}, summed
over different arrival times $t - \tau$.  For times longer than
$\tauN$, essentially all attached chains have fully adsorbed and a
monolayer of flattened chains starts to develop, which at times of
order $\tsatchem$ has almost covered the surface.

Up to $\tsatchem$ the fraction $f$ of bound monomers is approximately
the same for all chains and equal to $\omega$.  Thus $P(f)$ during
this stage is sharply peaked at $ \omega$.  In reality, we expect two
types of effects will somewhat broaden this distribution. The first is
fluctuations in $f$ values around $\omega$ due to random events
typical of multiplicative random processes characterizing
irreversibility (\eg monomers tra\-pped in knots might not be able to
bound to the surface).  Such fluctuations would be interesting to
characterize numerically following the example of ref.
\cite{konstadinidis:irrev_ads_monte_carlo} in which a monte-carlo
method was used to study the structure of a single fully collapsed
chain on a surface and an $\omega$ value was extracted using
$\gammabound \twid \omega N$ for large $N$.  We anticipate a second
source of broadening for chains arriving after $\toverlap$.  These
chains will have $f$ values which decrease with time since some
monomers will be unable to bound due to the presence of earlier
arrivers.  This would lead to a continuously broadening spectrum of
$f$ values with increasing time.  In practice this overlap time is
often close to the saturation time (e.g. for $N = 1000$, $\toverlap =
.25 \tsatchem$) so this broadening effect has little time to develop.

\begin{figure}
\centering
\resizebox{0.8\columnwidth}{!}{%
  \includegraphics{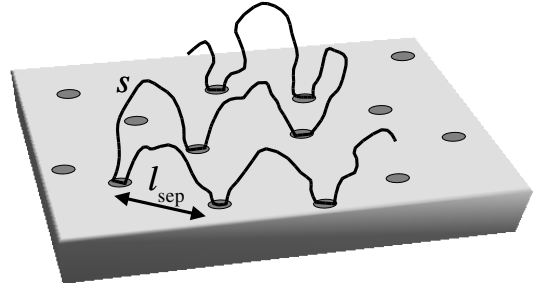} }
\caption{Typical configuration of adsorbed chain at late stages of
adsorption.  Such chains can adsorb only onto free empty sites (shown
as gray discs) which are separated by $\lsep$.  They thus form loops of $s$
monomers, with $as^{3/5}=\lsep$.}
\label{late_stages}      
\end{figure}

\subsection{Late Stages: Diffuse Outer Layer}

As the surface density approaches saturation, $a^2 \Gammabound \approx
1$, the availability of surface sites on the surface becomes scarce
and the late coming chains cannot fully adsorb on the surface as the
early arrivers did.  Let us suppose that the establishment of one
surface attachment requires an empty spot large enough to accommodate
$\ncont$ bound monomers, where $\ncont$ is a small-scale
species-dependent number similarly to $\omega$.  The surface density
of these ``supersites,'' $\rhosuper$, is becoming smaller with time
and the mean separation, $\lsep$, between neighboring supersites
increases accordingly, $\lsep \approx \rhosuper^{-1/2}$.  Now in order
for late-coming chains to adsorb onto these surface spots, they have
to form loops joining up these sites as shown in
fig. \ref{late_stages}.  We model the adsorption of chains at these
late stages by assuming that the size $s$ of such loops is the
equilibrium subcoil size corresponding to $\lsep$, \ie
                                                \begin{eq}{lsep}
a s^{3/5} = \lsep = \rhosuper^{-1/2} \period
                                                                \end{eq}
Thus chains which adsorb at the instant when the typical loop size in
eq. \eqref{lsep} is $s$, have a fraction of bound monomers given by 
                             \begin{eq}{chocolate}
f = {\partial \Gammabound \over \partial \Gamma}
                         = {\ncont \over s} \period 
                                                                \end{eq}

Now as $\Gammabound$ approaches its asymptotic value, $\Gammaboundinf$,
which is another nonuniversal small-scale dependent quantity of order $a^{-2}$,
then
                                                    \begin{eq}{coffee}
\Delta\Gammabound = \ncont \rhosuper = \ncont {a^{-2} \over s^{6/5}}
\comma\gap
                                                            \end{eq}
where $\Delta\Gammabound \equiv \Gammaboundinf - \Gammabound$ and we
used eq. \eqref{lsep}.  Eq. \eqref{coffee} is simpler to understand
starting from the completely saturated surface: it states that if one
were to unpeel chains from a completely saturated surface so as to create
$\rhosuper$ supersites, the number of bound monomers freed would be
$\ncont$ per supersite.

Now integrating eq. \eqref{chocolate} after using eq. \eqref{coffee}
one has
                                                \begin{eq}{output}
a^2 \Delta\Gammabound = \ncont (a^2 \Delta\Gamma/6)^6\comma
                                                                \end{eq}
which together with eq. \eqref{bad-good} of the early stages describe
the full theoretical relation between bound and total fractions plotted
in fig. \ref{granick_prl}(a).   Thus eq. \eqref{output} predicts a
very sharp saturation of the bound fraction as $\Gamma \gt \Gammainf$.

Now since in our picture every point along the
$\Gammabound(\Gamma)$ curve corresponds to a unique value of $f$,
see eq. \eqref{chocolate}, then the weighting of $f$ values is given
by
                                                \begin{eqarray}{dirty}
P(f) &=& \inverse{\Gammainf}
 \int d\Gamma\ \delta(f- \partial \Gammabound / \partial \Gamma)
                                \drop
 &=& \inverse{\Gammainf} 
\paren{\left|
\partial^2 \Gammabound/\partial\Gamma^2
\right|
_{\partial\Gammabound/\partial\Gamma=f}}^{-1}
 \comma
                                                                \end{eqarray}
which after using eq. \eqref{output}, leads to
                                                \begin{eq}{cow}
P(f) = Cf^{-4/5}\gap (f\ll 1)
                                                                \end{eq}
with $C\approx 6/(5 a^2\Gammainf [\ncont]^{1/5})$. 

\begin{figure}
\centering
\resizebox{0.9\columnwidth}{!}{
\includegraphics{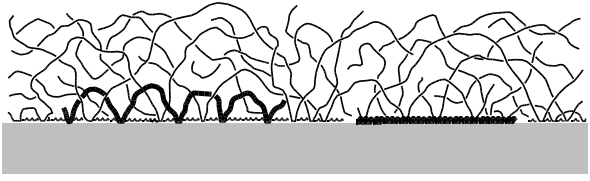}
}
\caption{Sketch of predicted final layer structure resulting from
irreversible polymer adsorption.  The layer consists of two parts (one
chain from each part is highlighted): (i) An inner region of flattened
down chains making $\omega N$ contacts per chain, where $\omega$ is of
order unity. (ii) A diffuse outer layer build up from chains each making
$f N \ll N$ contacts with the surface.  The values of $f$ follow
a distribution $P(f) \twid f^{-4/5}$.  Each $f$ value corresponds to 
a characteristic loop
size for a given chain, $s \approx \ncont/f$.  }
\label{layer}      
\end{figure}

The full theoretical prediction for the final layer's distribution of
$f$ values is thus the sum of eq. \eqref{cow} and a sharply peaked
function at $\omega$ from the early stages.  The overall distribution
is shown in fig. \ref{granick_prl}(b), binned into bins of width
$\Delta f =0.02$.  The binned distribution exhibits 2 peaks, of
different origin: the peak at large $f$ corresponds to early arriving
chains and its position is species-dependent while the peak at $f = 0$
represents a diffuse outer layer shown in fig. \ref{layer} and is due
to the universal small $f$ form, $P(f) \twid f^{-4/5}$.

Let us consider now the distribution of loop sizes in the resulting
diffuse layer, $\Omega(s)$, equal to  the number of loops of
length $s$ per unit area per unit loop length.  This may be found by
noting that the bound mass corresponding to a certain $d \Gammabound$
has a unique $s$ value and must thus equal the bound mass in loops of
length $s$, namely $\ncont \Omega(s) ds$.  Using eq. \eqref{coffee}
this leads to
                                                \begin{eq}{input}
\Omega(s) \approx a^{-2} s^{-11/5} \period
                                                                \end{eq}
What density profile $c(z)$ does this loop distribution generate as a
function of distance $z$ from the surface?  To determine the density
profile we follow similar arguments to those of refs.
\cite{gennes:ads_pol_loops,aubouy:scaling_flat_layers,semenovjoanny:loops_tails_europhys}
which relate loop distributions to density profiles in both specific
and general cases.  We do that by assuming that the density at a given
height $z$ is due to contributions from loops longer than $\sigma(z)$,
where $\sigma$ is the loop length which extends spatially up to height
$z$.  The density at a given height is determined by the number of
loops which are long enough to reach this height:
                                                \begin{eq}{printer}
c(z)  \approx {d\sigma(z) \over dz}  \int_{\sigma(z)}^{\infty}
ds\Omega(s)   
\period
                                                                \end{eq}
Since in our case loops are not stretched one has $z \approx a
\sigma^{3/5}$.  Thus using eq. \eqref{input} in eq. \eqref{printer} one
obtains the algebraic profile,
                                                \begin{eq}{density}
c(z) \approx c(a) \power{a}{z}{4/3} \comma
                                                                \end{eq}
which interestingly has the same scaling form as de Gennes'
self-similar profile of equilibrium polymer layers
\cite{gennes:self_similar_profile}.

\begin{figure}
\centering
\resizebox{0.9\columnwidth}{!}{%
\includegraphics{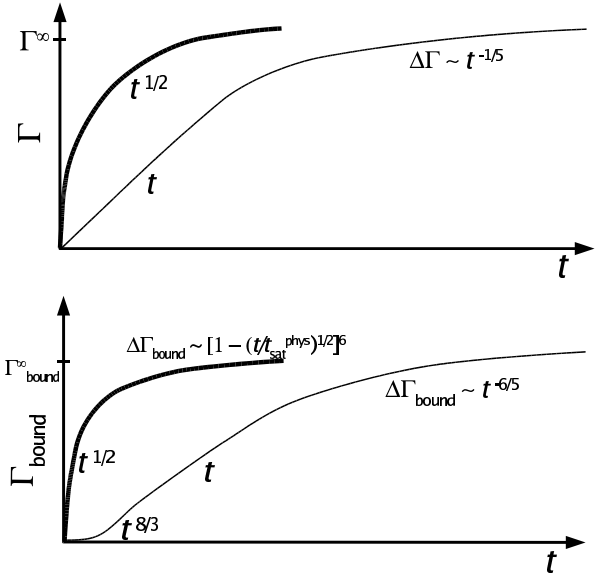}
}
\caption{Time evolutions of total, $\Gamma$, and surface bound,
$\Gammabound$, monomers per unit area, as a function of time, as
predicted by theory.  Both chemisorption (thin lines) and
physisorption (thick lines) are sketched.  The total mass for
chemisorption grows initially linearly, $\Gamma \twid t$, and then
slows down as the surface saturates, $\Delta \Gamma \equiv \Gammainf -
\Gamma \twid t^{-1/5}$.  For physisorption $\Gamma \twid t^{1/2}$ has
diffusion-controlled form for all times until saturation.  The surface
bound part for chemisorption undergoes three regimes: initially
$\Gammabound \twid t^{8/3}$, crossing over to $\Gammabound \twid t$
and then to the late stage saturation behavior, $\Delta \Gammabound
\twid t^{-6/5}$.  The surface bound part for physisorption follows
initially diffusion-controlled kinetics, $\Gammabound \twid t^{1/2}$,
but for longer times the kinetics slow down, $\Delta \Gammabound
\equiv \Gammaboundinf - \Gammabound \twid [1 -
(t/\tsatphys)^{1/2}]^6$.}
\label{gamma_kinetics}      
\end{figure}

Let us finally consider the {\em kinetics} of the building up of the
diffuse layer.  Since the rate of attachment is proportional to the
density of available surface sites, the early kinetics of
eq. \eqref{key} generalize to
                                                \begin{eq}{spray}
{d \Gamma \over dt} = \phibulk Q \Delta\Gammabound \period
                                                                \end{eq}
(In eq. \eqref{spray} we did not include free energy barriers due to
loops of the already partially formed layer which would present
exclude volume repulsion to new chains arriving at an empty
supersite. In fact, in order for our picture to be self-consistent
such effects must very small; were there any dangling loops near an
empty supersite they would adsorb onto this site.) Thus from
eq. \eqref{spray} one has
                                               \begin{eqarray}{like}
&a^2& \Delta \Gamma \approx F \, (\tsatchem/t)^{1/5}\comma\ \drop
&a^2& \Delta \Gammabound \approx G \, (\tsatchem/t)^{6/5}
                                \comma\ \ \ \  (t>\tsatchem) 
                                                                \end{eqarray}
after using eq. \eqref{output}.  The prefactors
$F=6(5\ncont/6)^{-1/5}$, $G=\ncont (C/6)^6$ are close to unity.
Together with the short time kinetics, eqs. \eqref{key} and
\eqref{not-good}, these evolutions are sketched in
fig. \ref{gamma_kinetics}.

So far in this section good solvent conditions were assumed.
Generalizing to the case of theta solvents is straightforward, by
replacing $\nu=3/5 \gt 1/2$.  One finds qualitatively similar results
with $\Delta \Gammabound \twid e^{-\Gamma a^2}$ replacing
eq. \eqref{output}, and the distribution of bound fractions is now
$P(f) \twid f^{-1}$, replacing eq. \eqref{cow}.


\section{Irreversible Physisorption}

We consider now the other important class of irreversible adsorption:
strong physisorption.  In this case monomer-surface bonds form
immediately upon contact and adsorption kinetics are
diffusion-controlled.  Despite the completely different kinetics
compared to chemisorption, we find the resulting polymer layers have
nevertheless almost identical structure.

\subsection{Early Stages: Monolayer Formation}

Initially the surface is empty and it is thus inevitable that any
chain whose center of gravity diffuses within the coil size
\cite{gennes:book,doiedwards:book} $\RF=aN^{3/5}$ of the surface
will adsorb.  Thus attachment of chains is diffusion-controlled and
the surface coverage grows as
                                                \begin{eq}{soup}
a^2 \Gamma(t)\approx {\phibulk \over a}\, (Dt)^{1/2} \comma
                                                                \end{eq}
where $D$ is the chain center of gravity diffusivity.  

Consider now the part of adsorbed mass which has bonded with the
surface, $\Gammabound$.  Immediately after the first attachment, the
subsequent monomer arrivals on the surface occur at the rate of their
diffusion on the surface.  Imagine there were no reactions and that
the surface was a penetrable plane.  Then, given the coil is initially
next to the surface, all monomers would cross the plane at least once
within the bulk coil relaxation time $\taubulk$
\cite{gennes:book,doiedwards:book}.  With reactions turned on, each
time a monomer reaches the plane, it would react with it.  We expect
the effect of the resulting constraint to further accelerate the rate
of new monomer arrival onto the surface and thus the chain
physisorption time $\tauads$ would be at most $\taubulk$.  We do not
attempt to analyze the details of these kinetics involving polymer
hydrodynamics with increasing number of constraints.  Such kinetics
would anyway be very difficult to detect experimentally since typical
coil relaxation times in solution are microseconds.  Here, it is
enough to know that $\tauads \lsim \taubulk$, which is supported by
numerical simulations in the absence of hydrodynamic interactions in
ref. \cite{shaffer:strong_ads_heteropolymers}.  Thus provided the
solution is dilute, $\phi<\phistar$, each chain adsorbs fast enough
onto the surface before other chains interfere with it.

Since the single-chain adsorption time is faster than new chain
arrival on the surface, a monolayer of flattened chains starts to
develop on the surface, just as in the case of chemisorption.
Defining $\omega$ to be the surface-bound part of a completely
collapsed chain, as in chemisorption,  we thus have
                                                \begin{eq}{hair}
\Gammabound = \omega \Gamma \comma \ \ 
                                                                \end{eq}
This is the initial linear regime in fig. \ref{granick_prl}(a). 

The kinetics of eq. \eqref{soup} proceed up to a time at which the
density of available surface sites starts to become small, \ie when $\Gamma
\approx a^{-2}$.  This occurs at a time of order 
                               \begin{eq}{spray-two}
\tsatphys \approx \taubulk \ \power{\phistar}{\phibulk}{2}\, N^{2/5} \period
                                                                \end{eq}
After this time, new chains must form loops joining disconnected empty
sites and a diffuse layer starts to build up.

\subsection{Late Stages: Diffuse Outer Layer}

For times longer than $\tsatphys$, the late coming chains start
to see a continuously decreasing density of available sites for
adsorption.  Will the adsorption kinetics continue to be
diffusion-controlled?  Consider a chain which was brought by diffusion
to within one coil size $\RF$ of the surface.  Let us see if a bond
forms before it diffuses away.  The number of collisions the chain
makes with a certain site on the wall within $\RF$ of its center of
gravity before it diffuses away is:
                                                \begin{eq}{collisions}
\Ncollisions \approx  (\taubulk / \ta)\, \Nsite
\comma \ \ \ \
\Nsite \approx r {N a^3 \over \RF^3} \period
                                                                \end{eq}
Here $\ta$ is the monomer relaxation time and $\Nsite$ is the mean
number of the coil's monomers per surface site within the coil's
projected surface area.  This would simply be of order $N a^3 /\RF^3$,
but the presence of the hard wall reduces it by a factor $r \equiv
\Zsurf(N)/ \Zbulk(N) = 1/N$ as in eq. \eqref{phisurf}.  Using
\cite{gennes:book,doiedwards:book} $\taubulk \approx \ta (\RF/a)^3$
one has from eq. \eqref{collisions}
                                                \begin{eq}{rain}
\Ncollisions \approx 1
                                                                \end{eq}
Thus the chain makes of order one surface contact with every site on
the surface within the coil size.  It follows that unless the surface
density of available sites is so small as to be of order one per
$\RF^2$, it is inevitable that during the chain's residence time on
the surface at least one monomer-surface bond will form.  Thus the
diffusion - controlled kinetics of eq. \eqref{soup} continue up to the
point where essentially every empty site is filled.

What is the mode of adsorption of these late coming chains which
attach to the few empty sites?  We model adsorption in these late
stages in a similar way to the build up of the diffuse outer layer in
chemisorption (section 4.2).  The late coming chains, after adsorbing
on an empty site large enough to accommodate $\ncont$ monomers, form
bridges to nearby empty sites which are loops of $s$ monomers when the
average separation between neighboring empty sites is $\lsep \approx a
s^{\nu}$ (see fig. \ref{late_stages}).  This separation becomes larger
as more and more chains adsorb.  Thus the diffuse layer density
profile $c(z)$, distributions of loop sizes, $\Omega(s)$, and
distribution of fraction of adsorbed monomers per chain, $P(f)$, are
the same as for chemisorption (section 3).  Including the monolayer
contribution due to short times, the resultant overall
$\Gammabound(\Gamma)$ and $P(f)$ relations are plotted in
fig. \ref{granick_prl}.  This is a repeat of the chemisorption curves
with the exception of the early $\Gammabound \twid \Gamma^{8/3}$
regime exclusive to chemisorption.  The sketch of the resulting
polymer layer is the same as fig. \ref{layer}.

The major distinction between chemisorption and physisorption arises
in the kinetics of layer build up.  The difference in the $\Gamma(t)$
kinetics is evident by comparing eq. \eqref{soup} for physisorption,
valid up to complete surface saturation, to eqs. \eqref{key} and
\eqref{like} for chemisorption.  Similarly, the kinetics of the bound
fraction, $\Gammabound$ are also very different.  For physisorption,
the short-time kinetics are given by substituting eq. \eqref{soup} in
eq. \eqref{hair} while the long-time behavior is found using
eq. \eqref{soup} in eq. \eqref{output}, leading to
                                                \begin{eqarray}{land}
&& a^2 \Delta \Gammabound 
\approx \drop
&& \casesbracketsshortii
        {\omega \phibulk a^{-1} (D t)^{1/2}}
                                {t \ll \tfinalphys}
        {\ncont \, (a^2 \Gammainf)^6\square{ 1- (t/\tfinalphys)^{1/2}}^6} 
                                {t\gt\tfinalphys}
                                                               \end{eqarray}
Here $\tfinalphys$ is the complete surface saturation time, after
which the diffuse layer has completely formed.  This timescale is of
the same order of magnitude as $\tsatphys$ since the surface coverage
of the outer layer, $\int_a^{\RF} dz\, c(z) \approx a^{-2}$, is of the
same order as the monolayer coverage.  Thus from eq. \eqref{soup} one
sees that the time for diffuse layer formation is of the same order as
the time of monolayer formation since both require the diffusion to
the surface of about the same quantity of mass.  The complete
evolution of $\Gamma(t)$ and $\Gammabound(t)$ are sketched in
fig. \ref{gamma_kinetics}.


\section{Discussion}

\subsection{Comparison of Theory with Experiment}

We studied theoretically the structure of polymer layers formed by
irreversible adsorption onto surfaces from dilute solution under good
or theta solvent conditions.  Relatively few theoretical works have
addressed irreversibility in polymer adsorption.  In refs.
\cite{barford:irrev_ads_1,barfordball:irrev_ads_2} irreversible
physisorption was studied analytically and numerically within the
framework of self-consistent mean field theory.  These workers found
that compared to equilibrium, the irreversibly formed layers are
different in that (i) the asymptotic surface coverage $\Gammainf$ was
larger in equilibrium, and (ii) in the irreversible case the density
profile $c(z)$ was found to be larger for small and large $z$ but
smaller for intermediate values as compared to equilibrium.  In
another work \cite{shafferchakraborty:pmma_chemisorption_kinetics}
single chain chemisorption of PMMA onto Al was studied by solving
numerically the loop kinetics, eq. \eqref{loops}, using chain
statistics corresponding to theta-solvent solutions.  It was found
that the chain adsorbs in a zipping mechanism the origin of which was
greatly enhanced reactivities for monomers neighboring a graft point.
These reactivities were taken as greatly enhanced on the basis of
electronic structure calculations
\cite{shaffer:reactions_metal_polymer_jcp,chakrabort:glassy_polymer_solid}.

In this work the cases of irreversible physisorption and chemisorption
were examined separately.  We found that in both cases the final layer
consists of two regions.  In an inner region of completely flattened
chains each chain has on average $\omega N$ monomers bound to the
surface.  The value of $\omega$ is species-dependent and represents
the effect of steric constraints at the monomer level which prevent
all monomers of the chain from bonding.  The outer region is a
tenuously attached diffuse layer of chains making $f N \ll N$ contacts
with the surface.  The distribution of $f$ values among chains is
universal and in a good solvent follows $P(f) \twid f^{-4/5}$ for $f
\ll 1$.  This double layer structure is illustrated in fig.
\ref{layer} which exhibits two peaks at small and large $f$,
respectively.  Our theory cannot capture the numerical value of the
peak amplitude near $f=\omega$ but the existence of a peak (or
possibly of a depletion region) is due to small-scale effects. The
first layer of flattened chains is special because these arrive at a
bare surface, and chain features on the scale of a monomer size are
involved.  In our picture, one can think of the next set of arriving
chains (which are just starting to build up the outer tenuously
attached region) as seeing a different surface, one containing a
certain areal fraction of empty sites less than unity. This set cannot
flatten down completely. The third set of arriving chains sees a
surface containing a yet lower density of free sites, and so on.  For
large layer numbers, a chain sees a universal surface whose features
evolve self-similarly with increasing layer number. This leads to a
universal power law form for $P(f)$. Returning to the earliest layers,
these are formed before universality onsets, reflected in a
non-universal feature in $P(f)$ near $f=\omega$.

Using infrared absorption spectroscopy, measurements of $P(f)$ for
irreversible physisorption of PMMA onto oxidized silicon have been
pioneered in the experiments of refs.
\cite{frantzgranick:pmma_bound_fraction_early,schneider:granick_bimodal,douglas:kinetics_pol_ads_jpc}
measurements from which are reproduced in fig. \ref{granick_prl}(b).
In this figure, the theoretical prediction is compared to experiment
by binning values in ranges $\Delta f=0.02$ and converting analytical
predictions to a histogram.  Since $f=0.08$ is the lowest observed
experimental value, we cut off the theoretical distribution at the
same point for the sake of comparison.  The contribution from the
inner layer is represented by a delta-function centered at $\omega
\approx 0.47$, though as discussed in section 4, we expect this to be
broadened.  We see that the agreement with experiment is good and
captures the bimodal shape of the distribution.  We note that
measuring lower $f$ values to see a clear signature of a power law
regime may be difficult: our theory suggests these chains are a small
fraction of the total chain population (the integral of $P(f)$ is
dominated by large $f$).  However since those chains lie in the
outermost region of the layer, they would determine important physical
properties such as hydrodynamic thickness and the strength of
interaction of the polymer layer with an approaching interface.  In
comparing with theory above, we have used our predictions for good
solvents.  However, the experiments of refs.
\cite{schneider:granick_bimodal,douglas:kinetics_pol_ads_jpc} were
performed under solvent conditions slightly better than theta. Thus it
may be that a more appropriate comparison is with our theory
specialized to theta solvents, for which the predicted $P(f)\twid 1/f$
is very similar.

The building up of the double layer is apparent in the shape of
$\Gammabound(\Gamma)$ where $\Gammabound$, $\Gamma$, are surface-bound
and total surface coverage, respectively.  For irreversible
physisorption we found that initially $\Gammabound \approx \omega
\Gamma$.  As the surface saturates, the diffuse layer starts to
develop and $(\Gammaboundinf - \Gammabound) \twid (\Gammainf -
\Gamma)^6$ as the asymptotic values (denoted by $\infty$) are
approached which are of order $a^{-2}$, where $a$ is monomer
size.  We thus predict a very sharp saturation of the bound fraction
as $\Gamma$ saturates; the resulting curve is plotted in
fig. \ref{granick_prl}(a).  A very similar curve has been measured in
the experiments discussed in the previous paragraph     
\cite{schneider:granick_bimodal,douglas:kinetics_pol_ads_jpc}, where
the initial slope was $\omega \approx 0.47$, the value we used in
fig. \ref{granick_prl}(b).  In addition the shape of the curve was
found to be independent of chain length, which is also consistent with
our model.  For chemisorption, the curve is similar but we predict an
additional $\Gammabound \twid \Gamma^{8/3}$ regime as shown in
fig. \ref{granick_prl}(a).

Contrary to physisorption, the experimental picture for chemisorption
is much less clear, though since time\-scales are intrinsically much
longer, the kinetics might be easier to probe.  Experiments on various
systems have been performed
\cite{lenk:functionalized_pmma_on_gold,%
schlenoff:functionalized_ps_on_gold,tsao:functionalized_pdms_on_gold,%
cosgrove:chemisorption_ftir,cosgrove:chemisorption_neutron_scattering,%
cosgrove:phys_and_chem_adsorption_langmuir} the results of which
however cannot be interpreted within the framework of our present
theory since they involved simultaneous physisorption and
chemisorption.  In certain cases the degree of polymer
functionalization was also varied
\cite{schlenoff:functionalized_ps_on_gold,tsao:functionalized_pdms_on_gold,%
lenk:functionalized_pmma_on_gold}.  Clearly, the dependence of resulting
polymer structures on degree of polymer or surface functionalization
is an interesting aspect deserving further study.

This paper has addressed both the structure of the final layer and the
irreversible kinetics of its formation.  Chemisorption kinetics are
very slow and are chemically controlled.  We found that initially
$\Gamma \twid t$, followed by $\Gamma \twid t^{-1/5}$ as the surface
saturates.  We remark that for long enough times, adsorption onto a
planar surface always becomes diffusion-controlled after a
$Q$-dependent timescale
\cite{ben:reactiface_fund_letter,ben:reactiface_fund}, where $Q$ is
reaction rate upon monomer-surface contact.  Here our assumption is
that $Q$ is small enough such that this cross-over time is very large,
which is the typical case for ordinary chemical species.  For
physisorption, diffusion controlled kinetics apply until saturation,
$\Gamma \twid t^{1/2}$.  The corresponding time dependencies for
$\Gammabound$ follow from $\Gammabound(\Gamma)$ and are shown in
fig. \ref{gamma_kinetics}.

A large part of this work has focused on single-chain chemisorption.
Unlike physisorption where single chain adsorption is complete at most
after the coil's bulk relaxation time, $\taubulk$, we found that the
adsorption time in chemisorption is much larger, $\tauads \approx
Q^{-1} N^{3/5}$.  Thus for typical $Q \approx 1 {\rm s}^{-1}$, $N =
1000$, one has $\tauads \approx 60$s, a time accessible to
experimental measurements.  During $\tauads$ the chain adsorbs onto
the surface in a mode we call accelerated zipping: initially the chain
adsorbs in a zipping mechanism growing outwards from the first
attachment point with the surface, but with increasing time distant
monomers adsorb forming large loops and new sources for further
zipping which accelerate the chain's collapse.  The distribution of
these loop sizes follows a power law $\omega(s) \twid s^{-7/5}$ while
the number of bound monomers grows as $\gammabound(t) \twid t^{5/3}$.

\subsection{Differences between Irreversible and Equilibrium Layers}

We conclude with a general comparison between the final
non-equilibrium layers predicted by our theory and equilibrium
layers. This discussion is limited to good solvents.  Our results for
the final loop distribution of the layer, $\Omega(s) \twid s^{-11/5}$
and density profile $c(z) \twid z^{-4/3}$ have interestingly identical
scaling form to the corresponding equilibrium results. On the other
hand, the configurations of individual chains are very different.  In
the equilibrium layer a given chain has $N D(s) $ loops of length $s$
or greater, where $D(s) \equiv \int_s^\infty ds' \Omega(s') \twid
s^{-6/5}$.  Because of screening effects these are essentially
independent blobs and their 2D spatial extent parallel to the surface
is $ [N D(s)]^{1/2} a s^{3/5} = a N^{1/2}$. This is true for all
scales $s$; in particular, there is order one loop of length
$N^{5/6}$, also of size $aN^{1/2}$. Hence a typical chain has a size
\cite{semenovjoanny:kinetics_adsorption_rouse} of order $a N^{1/2}$,
the ideal result (to within logarithmic corrections
\cite{semenovjoanny:kinetics_adsorption_rouse}).  Chains in the
non-equilibrium layer are by contrast of size $aN^{3/5}$, due to the
adsorption kinetics which in the case of chemisorption entailed
occasional adsorption of very large loops of length $N$ which ensured
the final state of the chain extends a distance of order $a N^{3/5}$
parallel to the surface.  We did not attempt to unravel the complex
details of physisorption, but our assumption was that the 3D coil ends
up being roughly projected onto the 2D surface, leading again to a
lateral size $a N^{3/5}$.

The most fundamental distinguishing feature of the non-equilibrium
layer is that different chains have different loop distributions and
fractions of adsorbed segments, $f$, which are frozen in time.  There
is no longer one single thermodynamic $f$.  Instead there are
infinitely many classes of chains, each with its own $f$ value, and
the number of chains in each class is proportional to $P(f)$. Each
chain has a characteristic loop size, $s\twid 1/f$.

This is very different to the equilibrium layer where all chains are
statistically identical, in that every chain explores in time the same
range of $f$ values with the same relative weighting $\Peq(f)$.  In
fact, one can show that for large $N$ this distribution $\Peq$ is very
sharply peaked at a mean value $\fbar$ of order unity, with width of
order $N^{-1/6}$. (The mean value is due to small loops of order unity
while the fluctuations are due to the mass in very large loops with
length of order $N^{5/6}$. Since there is order one such loop, this
diminishes $f$ by an amount of order $N^{-1/6}$.)  Thus even at any
fixed moment in time it is still a true statement that almost all
chains have the same $f$ value, $\fbar$, to within fluctuations which
are small for large enough $N$.  We remark that there is a small
population of chains, a fraction of order $N^{-1/5}$ of the total,
with $f$ values far removed from the mean: these are the chains with
tails \cite{semenovjoanny:loops_tails_europhys} of length of order $N$
which determine the layer height $\approx a N^{3/5}$.

These differences in chain statistics between irreversible and
equilibrium layers have important implications for various physical
properties of the layer. Consider a physisorbing system whose surface
relaxation kinetics are very slow, but not truly irreversible (we
expect this is a rather common case).  In this case for times of order
the layer formation time we expect the layer is well-described by the
irreversible $P(f)$.  For much longer timescales on which surface
bonds are reversible equilibration processes will slowly evolve $P(f)$
into $\Peq(f)$.  Our picture predicts that during this process the
total coverage, density profile, and loop distribution remain
unchanged to within prefactors of order unity.  Single chain
statistics will be significantly affected, however, and individual
chains must shrink in order to occupy a much smaller region in space.
This change is expected to profoundly modify the physical properties
of the layer.  For example, the fraction of chains with long loops and
tails of order $N$ will be reduced by a factor $N^{-1/5}$, thus
strengthening the outer region of the layer which is exposed to the
bulk.  Similarly in experiments probing the kinetics of exchange
between chains in the bulk and those in the layer
\cite{pefferkon:ads_pol_dynamics_bimolecular,pefferkon:ads_pol_exhange_kinetics,%
varoqui:mobility_ads_pol_review,%
frantzgranick:kinetics_adsorption_desorption_prl,%
frantzgranick:ps__cyclohexane_exchange_macro,
schneidergranick:kinetic_traps_exchange,
johnsongranick:exchange_kinetics_pmmma,%
frantzgranick:pmma_bound_fraction_early,
fusantore:peo_competitive_adsorption,fusantore:age_relaxation_peo_exchange,%
mubarekyansantore:N_age_peo_exchange,mubarekyansantore:barrier_peo_exchange},
we expect strong aging effects since the number of easily
exchangeable, loosely attached chains is decreasing with increasing
aging time.  This would lead to a decrease of the initial exchange
rate with aging time as observed
\cite{frantzgranick:ps__cyclohexane_exchange_macro,
schneidergranick:kinetic_traps_exchange,fusantore:age_relaxation_peo_exchange,%
mubarekyansantore:N_age_peo_exchange}.  Further theoretical work is
clearly needed to interpret the phenomenology of these experiments.
One unexplained observation, for example, concerns poly(ethylene
oxide) adsorption onto silica where the bulk-surface exchange rate of
chain subpopulations in non-equilibrium layers has been measured to be
independent of the time the subpopulation was incorporated in the
layer \cite{mubarekyansantore:barrier_peo_exchange}.

What is the origin of the theoretically predicted irreversible layer's
overall loop distribution being the same as that in equilibrium?  This
is rooted in the kinetics of our model.  In the late stages, we
assumed the equilibrium relation between loop size and number of
monomers $s$ in the loop, $R\twid s^{\nu}$.  That is, a given loop is
assumed to have equilibrium statistics.  It is important to note that
this does not by itself lead to the equilibrium loop distribution. Our
basic assumption on the late stage kinetics is that the kinetically
selected loop size $R$ at any moment is determined by the current
density of free supersites, $\rhosuper \twid 1/R^2$. Taken together
with $R\twid s^{\nu}$, this then leads to the $\Omega(s)$ of
eq. \eqref{input}. These kinetics are mean field in character,
assuming uniformly smeared free sites determined by the current global
value of $\rhosuper$ rather than any local feature. The correct choice
of kinetics is governed by the kinetics of the chain adsorption
process.

To illustrate this point, consider a general situation, involving
chain adsorption onto a $d$-dimensional surface.  Suppose the
mechanism of adsorption is pure zipping and we follow the process from
a starting situation where the surface has a uniform density of empty
surface sites, $\rhosuper = l^{-d}$.  Roughly speaking such a zipping
chain performs a simple random walk on the surface, each step filling
one empty site and producing a loop of length $s=l^{1/\nu}$.  The
statistics of this process depend on dimensionality. (i) Suppose
$d=1$.  Since a 1D random walker explores space
``compactly'' \cite{gennes:polreactionsi}, i.e.  visits each site
within its exploration volume many times, by the time the zipping is
complete and the chain has fully adsorbed a surface patch depleted of
empty sites will have been created in the region where the chain has
adsorbed.  Later-arriving chains can then occupy only the empty
regions between such patches created by previously adsorbed chains.
We expect the density of free sites to fluctuate very strongly,
invalidating any mean field approach. (ii) $d$=3.  The zipping process
is now a ``non-compact'' random walk, occupying only a small fraction
of the free surface sites within its exploration volume.  Fluctuations
are small, and the mean field approach which assumes an essentially
uniform distribution of free sites of density $\rhosuper$ decreasing
quasistatically in time is valid.  This is the assumption implicit in
the present paper.  (iii) $d$=2. Random walks are now marginally
compact and one expects logarithmic corrections to the mean field
picture.

The situation treated in this paper is of course $d=2$. Thus one might
expect logarithmic corrections to the loop distribution of eq.
\eqref{input} which was arrived at via the mean field argument.  In
fact, for two reasons we expect the zipping process to execute a
surface walk somewhat expanded from a simple marginally compact 2D
random walk.  Firstly, no site can be adsorbed onto twice by the
zipping random walk.  Thus walks are repelled from frequently visited
areas.  This is a kind of self-avoidance, and one anticipates swelling
of the walk somewhat similarly to the ``true'' self-avoiding walk
\cite{amit:true_saw} and the ``kinetic growth'' walk
\cite{pietronero:kinetic_growth_walk_prl}.  Secondly, the chain
adsorption processes we have studied are not simple zipping.  For
chemisorption, the zipping is accelerated by occasional very large
steps producing loops of all sizes up to $N$.  These large excursions
result in the flattened chain occupying an area of order $N^{6/5}$
rather than $N$. Since the fraction $N^{-1/5}$ of occupied sites
within this area is small, one returns to a non-compact situation
where mean field applies.  In reality, this is over-simplified since
local zipping from nucleation points presumably involves correlations
of the type characterizing pure zipping, so non-uniform depletion of
empty sites may occur near zipping centers.  For physisorption the
process is even more complex, presumably involving a combination of
local diffusion-mediated zipping and large loop formation, but one
expects a projection area onto the surface again of order $N^{6/5}$
which naively suggests non-compact character.

From the above remarks, it is clear that stepping beyond our rather
simple approach to irreversible adsorption involves many complex
issues. We hope our work motivates future experiment and theory in
this direction.

\begin{acknowledgement}

This work was supported by the National Science Foundation under grant
no. DMR-9816374.  

\end{acknowledgement}


{\appendix

\section{Single Chain Chemisorption:  Interior Monomers Attach First}

In this Appendix we show that a single chain is much more likely to
chemisorb on a surface by one of its interior monomers rather than
with one of its ends.  Consider the $s^{\rm th}$ monomer, where by
definition $s$ is the length of shorter part of the chain, the other
having length $N-s$.  The reaction rate of this monomer with the
surface is proportional to the partition function of the chain
anchored by this monomer on the surface:
                                                \begin{eqarray}{contact}
\Zcont (s) &=&\mu^N \eta(s/N) N^{\gamma_2 - 1} \comma
\drop
\eta(x) &\approx& \casesbracketsshortii
                {x^{\gamma_2 - \gamma_1}}    {x \ll 1}
                {1}                          {x \gt 1/2}
                                                                \end{eqarray}
where $\mu$ is a constant of order unity.  We derived $\Zcont (s)$ by
demanding that (i) it is a power in $s$ and (ii) in the limits $s \gt
1$ and $s \gt N/2$, one obtains the known partition functions
$\Zcont (1) \approx \mu^N N^{\gamma_1
-1}$, $\Zcont (N/2) \approx \mu^N N^{\gamma_2 -1}$, respectively.
The numerical values of $\gamma_1,\gamma_2$ are
\cite{duplantier:networks,semenovjoanny:kinetics_adsorption_rouse,%
debelllookman:surface_exponents_review} $\gamma_1 \approx 0.68$,
$\gamma_2 = \gamma -1 \approx 0.16$, where $\gamma$ is the
susceptibility exponent \cite{gennes:book}.

Now the total rate for any monomer to react is proportional to
$\int_1^{N/2} \,ds Z_{\rm cont}(s)$.  Since $\gamma_2-\gamma_1 > - 1$
this integral is dominated by large $s$.  Hence it is much more likely for a
monomer in the middle to be the first one to react with the surface,
rather than an end, roughly by a factor $N \Zcont(N) / \Zcont(1)
\approx N^{1 + \gamma_2-\gamma_1} \approx N^{0.48}$.  Thus the
accelerated zipping always propagates outwards from an interior
monomer, as schematically illustrated in fig. \ref{acc_zipping}.


\section{Proof that Eq. \eqref{loops} Satisfies Mass Conservation}

We show in this Appendix that eq. \eqref{loops} conserves the total number of
monomers $\Mtot \equiv \int_0^\infty ds \, s\, \omegatau(s)$, by showing
that $d \Mtot/d\tau = 0$.  This is equivalent to showing that the
integral of the rhs of eq. \eqref{loops} over all positive $s$ is
zero.  Changing variables from $s'$ to $x=s/s'$ in the first term, and
to $x=s'/s$ in the second term, on the rhs of eq. \eqref{loops},
respectively, one has
                                                \begin{eqarray}{coin}
&&2 \int_{s}^\infty ds'\, \omegatau(s')\, k(s|s') 
 - \int_0^s ds'\, \omegatau(s)\, k(s'|s) =
                                                \ddrop
&& {2 \over s^{\nu}} \int_0^1 dx\, \omegatau \paren{s\over x} 
{\zeta(x) \over x^{1-\nu }}
- {\omegatau(s) \over s^{\nu}} \int_0^1 dx\, \zeta(x)
                                                                \end{eqarray}
after using eq. \eqref{superdil}.  Multiplying eq. \eqref{coin} by
$s$, integrating over all positive $s$ and changing variables from $s$
to $y=s/x$, one finds that the result vanishes by virtue of the
identity $2 \int_0^1 dx \, x \zeta(x) = \int_0^1 dx \, \zeta(x)$ for $\zeta(x)$
symmetric around $x=1/2$.  Thus $\Mtot$ is conserved.


\section{Derivation of Eq. \eqref{mother} from Eq. \eqref{rum}}

Defining $\xi \equiv N -s$ and $\omegatwiddle_{\tau}(\xi) \equiv
\omegatau(N-\xi)$, one has from eq. \eqref{rum}
                                                \begin{eq}{xi}
{d \over d \tau} \int_0^\xi d\xi'\,\omegatwiddle_{\tau}(\xi') 
=
- \int_0^\xi d \xi'{A\,  Q\,  \omegatwiddle_{\tau}(\xi') \over (\xi-\xi')^{\nu}}
\period
                                                                \end{eq}
Laplace transforming $\xi \gt E$ (we allow $\xi$ to take any positive
value, \ie we seek the $N \gt \infty$ solution of eq. \eqref{rum})
one has from eq. \eqref{xi}:
                                                \begin{eq}{roro}
{d \over d \tau} \omegatwiddle_{\tau}(E) = - A\, Q\,  E^{\nu} \omegatwiddle_{\tau}(E)
\comma
                                                                \end{eq}
which after integration over $\tau$ becomes
                                                \begin{eq}{riri}
\omegatwiddle_{\tau}(E) = e^{-A\, Q\,  \tau\,  E^{\nu}}
\period
                                                                \end{eq}
We used the initial condition $\omegatwiddle_0(E) = 1$, since
$\omega_0(\xi) = \delta(\xi)$ (see eq. \eqref{rum}).  Laplace
inverting \cite{feller:prob2} eq. \eqref{riri} one
recovers eq. \eqref{mother}.


\section{Steady State Solution of Eq. \eqref{deq} }

We seek a quasi-static power law solution, $\omegatau(s) =
H/s^\alpha$, to eq. \eqref{small-one}.  Substituting to the left
hand side of eq. \eqref{small-one} one has
                                                \begin{eqarray}{cycles}
&& \int_s^{\infty} ds' \, \lambda(s|s') \, \omegatau(s') = \drop
&& - {Q H \over s^{\alpha+\nu-1}}\,
\int_0^1 dy \, y^{\nu - 2 + \alpha} \int_{y}^{1-y}dx \zeta(x) 
\comma
                                                                \end{eqarray}
after changing variables to $y = s/s'$.  Here we took the well-defined
limit $N \gt \infty$ since the solution we seek should be independent
of the tail's length.  The only value of $\alpha$ for which
eq. \eqref{cycles} is zero is $\alpha =  2 - \nu$.
The self-consistency of the quasi-static solution, eq. \eqref{good},
can be verified by substitution in eq. \eqref{deq} and showing that
for loop sizes where the solution is valid ($s <\staumax$) corrections
arising from integration over $s'> \staumax$ are small.



}



\end{document}